\documentclass[twocolumn,trackchanges]{aastex631}
\usepackage{amsmath,booktabs,amssymb}
\usepackage{multirow}
\usepackage{chemformula}
\usepackage[version=4]{mhchem}
\usepackage{tabularx}
\usepackage{threeparttable}
\usepackage{booktabs}
\usepackage[version=4]{mhchem}
\usepackage{subfigure}
\usepackage{comment}
\usepackage{tablefootnote}
\begin{document}

\title{Rotational Spectroscopy and Tentative Interstellar Detection of 3-Hydroxypropanal (\ce{HOCH2CH2CHO}) in the G+0.693-0.027 Molecular Cloud} 

\author[0000-0001-5020-5774]{Zachary T. P. Fried}
\affiliation{Department of Chemistry, Massachusetts Institute of Technology, Cambridge, MA 02139, USA}
\author[0000-0002-3849-1533]{Roman A. Motiyenko}
\affiliation{Univ. Lille, CNRS, UMR 8523 - PhLAM - Physique des Lasers Atomes et Mol\'{e}cules, F-59000 Lille, France}
\author[0000-0001-9629-0257]{Miguel Sanz-Novo}
\affiliation{Centro de Astrobiología (CAB), INTA-CSIC, Carretera de Ajalvir km 4, Torrejón de Ardoz, 28850 Madrid, Spain}
\author[0000-0003-3567-0349]{Lucie Kolesnikov\'{a}}
\affiliation{Department of Analytical Chemistry, University of Chemistry and Technology,
Technick\'{a} 5, 166 28 Prague 6, Czechia}
\author [0000-0002-2929-057X]{Jean-Claude Guillemin}
\affiliation{ Univ Rennes, Ecole Nationale Supérieure de Chimie de Rennes, CNRS, ISCR – UMR6226, F-35000 Rennes, France}
\author[0000-0003-0629-8277]{Laurent Margul\`es}
\affiliation{Univ. Lille, CNRS, UMR 8523 - PhLAM - Physique des Lasers Atomes et Mol\'{e}cules, F-59000 Lille, France}
\author[0000-0002-6220-2092]{Tereza~Uhl\'ikov\'a}
\affiliation{Department of Analytical Chemistry, University of Chemistry and Technology,
Technick\'{a} 5, 166 28 Prague 6, Czechia}
\author[0000-0003-0046-6217]{Arnaud Belloche}
\affiliation{Max-Planck-Institut f\"{u}r Radioastronomie, Auf dem H\"{u}gel 69, 53121 Bonn, Germany}
\author[0000-0001-9133-8047]{Jes K. J{\o}rgensen}
\affiliation{Centre for Star and Planet Formation, Niels Bohr Institute \& Natural History Museum of Denmark, University of Copenhagen,
\O{}ster Voldgade 5–7, 1350 Copenhagen K., Denmark}
\author [0009-0008-1171-278X]{Martin S. Holdren}
\affiliation{Department of Chemistry, Massachusetts Institute of Technology, Cambridge, MA 02139, USA}
\author[0000-0003-2760-2119]{Ci Xue}
\affiliation{Department of Chemistry, Massachusetts Institute of Technology, Cambridge, MA 02139, USA}
\author[0000-0002-4566-4930]{\v{S}t\v{e}p\'{a}n~Urban}
\affiliation{Department of Analytical Chemistry, University of Chemistry and Technology,
Technick\'{a} 5, 166 28 Prague 6, Czechia}
\author[0000-0003-4493-8714]{Izaskun Jiménez-Serra}
\affiliation{Centro de Astrobiología (CAB), INTA-CSIC, Carretera de Ajalvir km 4, Torrejón de Ardoz, 28850 Madrid, Spain}
\author[0000-0002-2887-5859]{Victor M. Rivilla}
\affiliation{Centro de Astrobiología (CAB), INTA-CSIC, Carretera de Ajalvir km 4, Torrejón de Ardoz, 28850 Madrid, Spain}
\author[0000-0003-1254-4817]{Brett A. McGuire}
\affiliation{Department of Chemistry, Massachusetts Institute of Technology, Cambridge, MA 02139, USA}
\affiliation{National Radio Astronomy Observatory, Charlottesville, VA 22903, USA}

\correspondingauthor{Zachary T. P. Fried, Brett A. McGuire}
\email{zfried@mit.edu, brettmc@mit.edu}

\begin{abstract}
We synthesized the astrochemically relevant molecule 3-hydroxypropanal (\ce{HOCH2CH2CHO}) and subsequently measured and analyzed its rotational spectrum in several frequency regions ranging from 130 to 485 GHz.  We analyzed the ground vibrational state as well as the two perturbed lowest-lying vibrationally excited states. With the resulting rotational parameters, we searched for this molecule in the Sagittarius B2(N) and NGC 6334I hot cores, the IRAS 16293-2422B hot corino, and the G+0.693-0.027 and TMC-1 molecular clouds. Rotational emission of 3-hydroxypropanal was tentatively detected toward G+0.693-0.027 and a column density of (8.6$\pm$1.4)$\times$10$^{12}$ cm$^{-2}$ was determined. However, this molecule was not detected in the other sources that were investigated. The chemical implications of this tentative discovery are analyzed and several potential chemical formation pathways of this species are discussed.
\end{abstract}

\section{Introduction}

\label{sec:intro}
Recently, the \ce{C3H6O2} isomers have been of particular astrochemical interest. Three of these isomers, namely ethyl formate \citep{ethyl_formate,rivilla_ethylformate_2017,peng_alma_2019}, methyl acetate \citep{methyl_acetate}, and hydroxyacetone \citep{hydroxyacetone}, have been detected toward various interstellar sources. That being said, the most thermodynamically stable isomer propanoic acid was not detected in searches toward Orion KL and Sagittarius (Sgr) B2 \citep{ilyushin_submillimeter_2021,2025A&A...698A.143B}, thus suggesting that the gas-phase interstellar abundances do not solely follow trends of thermodynamic stability -- a phenomenon observed in several other isomeric families (e.g. \citealt{loomis_investigating_2015,shingledecker_case_2019, mininni_guapos_2020,agundez_chemistry_2023,rivilla_first_2023,san_andres_first_2024}). The higher-energy isomers methoxyacetaldehyde \citep{methoxyacetaldehyde} and lactaldehyde (also known as 2-hydroxypropanal) \citep{lactaldehyde} were also not detected in searches toward these same star-forming regions. 

An additional \ce{C3H6O2} isomer of astrochemical interest is 3-hydroxypropanal (\ce{HOCH2CH2CHO}). This species has been recommended as a strong detection target by several groups (e.g. \citealt{wang23}, \citealt{fried}). Notably, \citet{wang23} recently investigated the formation of various \ce{C3H6O2} isomers from methanol-acetaldehyde ices that were irradiated with energetic electrons at low temperatures, simulating molecular cloud conditions. They ultimately observed the formation of hydroxyacetone and methyl acetate (which were both detected in the interstellar medium) along with 3-hydroxypropanal. They note that 3-hydroxypropanal is formed via the recombination of the \ce{CH2OH} and \ce{CH2CHO} radicals, which may be hydrogen loss products of methanol and acetaldehyde: 

\begin{equation} \label{eq:radical_recomb}
        \ce{CH2OH + CH2CHO -> HOCH2CH2CHO}
\end{equation}

Prop-1-ene-1,3-diol, which is the enol tautomer of 3-hydroxypropanal, was also observed. \citet{wang23} proposed that this species forms from keto-enol tautomerization of 3-hydroxypropanal, which has been shown to be possible in irradiated ice experiments that mimic the interaction of cosmic rays with interstellar ices under molecular cloud conditions \citep{kleimeier_interstellar_2021}.

Additionally, \citet{wang_interstellar_2024} also observed that 3-hydroxypropanal was formed (along with lactaldehyde, ethyl formate, and prop-1-ene-1,3-diol) in irradiated CO-ethanol ices via Reaction~\ref{eq:radical_recomb2}:

\begin{equation} \label{eq:radical_recomb2}
        \ce{HCO + CH2CH2OH -> HOCH2CH2CHO}
\end{equation}

Moreover, 3-hydroxypropenal (\ce{HOCHCHCHO}) was detected toward IRAS 16293-2422B \citep{hydroxypropenal,muller_rotation-tunneling_2024}. This molecule is structurally similar to 3-hydroxypropanal, with the only difference being a single extra degree of unsaturation. The formation pathway of this molecule is also proposed to have commonalities with 3-hydroxypropanal, as they both stem from reactions with the \ce{CH2CHO} radical \citep{hydroxypropenal}. This further suggests that 3-hydroxypropanal may be generated in this source.

In order to detect this molecule in radio astronomical observations, we must first measure and analyze its rotational spectrum. For detection in the star-forming regions in which other \ce{C3H6O2} isomers have been observed, an analysis of the millimeter and sub-millimeter wave spectrum is necessary.  The warm temperatures of these regions, possibly reaching several hundred Kelvin, cause the most intense rotational transitions to occur within these frequency ranges. That being said, a rotational investigation of this molecule had not been previously conducted. Therefore, in this work, we synthesized 3-hydroxypropanal and subsequently collected and fitted its millimeter and sub-millimeter wave rotational spectrum. This rotational data then enabled a search for the molecule in the Sgr B2(N) and NGC 6334I hot cores, the IRAS 16293-2422B hot corino, and the G+0.693-0.027 and TMC-1 molecular clouds. Section~\ref{sec:calc} outlines the quantum chemical calculations performed to support the experiments. Sections~\ref{sec:exp_description} and \ref{sec:exp_results} then detail the experimental procedure to synthesize the molecule along with the spectroscopic experiments and results. Finally, in Section~\ref{sec:obs_results}, we present the observational search for 3-hydroxypropanal in the various interstellar sources and discuss some potential chemical formation pathways of this species.  

\section{Quantum Chemical Calculations}

\begin{figure}[h!]

\begin{center}
\includegraphics[width=\columnwidth]{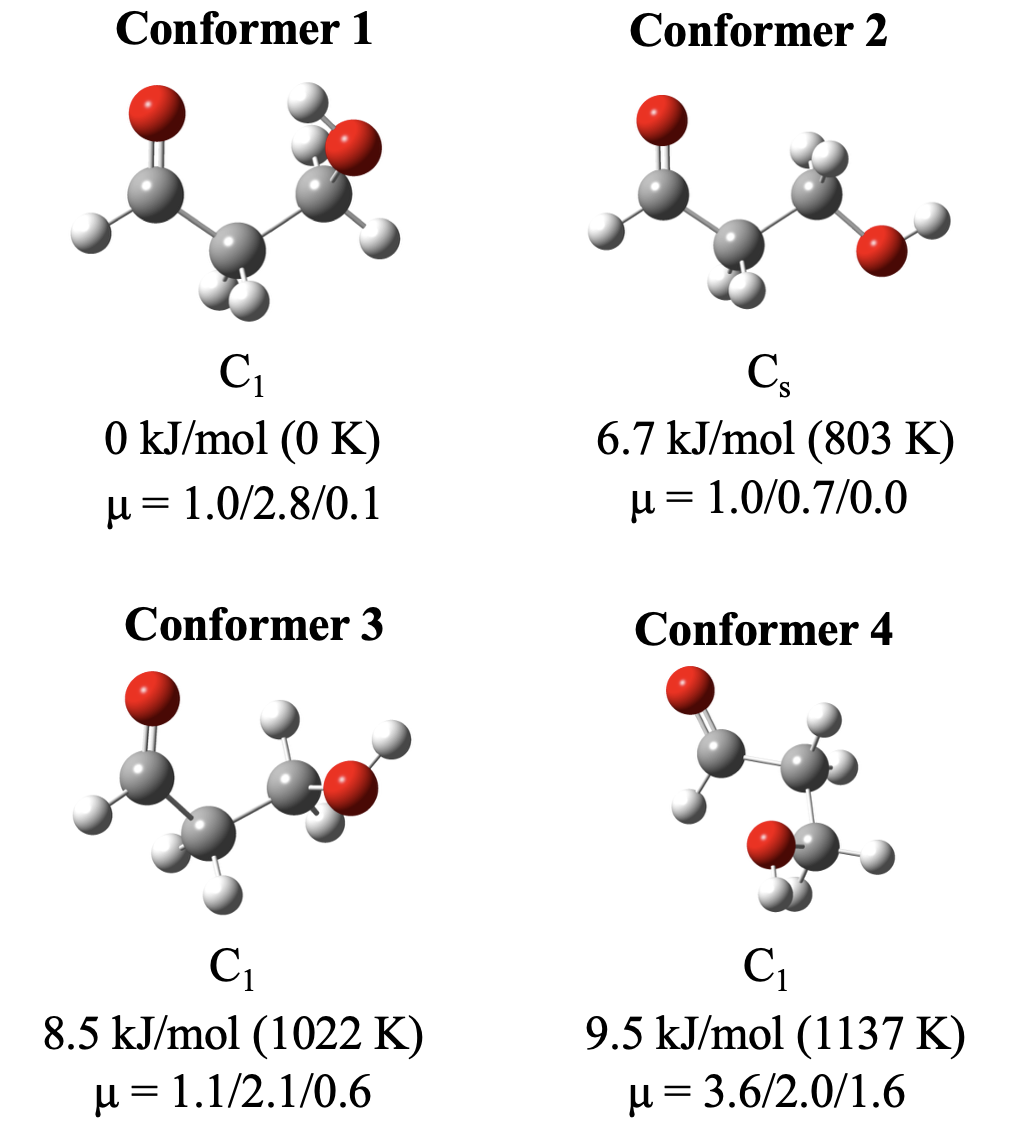}
\caption{
The structures and relative zero-point-corrected energies (in kJ/mol and kT-equivalent temperature) of the four most stable conformers of 3-hydroxypropanal. The dipole moments are listed as $\mu_a$/$\mu_b$/$\mu_c$. The point group of each conformer is also shown. See text for computational details. The red spheres denote oxygen atoms, the grey spheres are carbon atoms, and the white spheres depict hydrogen atoms.}
\label{fig:conformers}
\end{center}
\end{figure}

The conformational space of 3-hydroxypropanal is potentially vast due to the three single-bond rotations in its backbone. 
In order to determine which conformers would be present in our experimental sample, we conducted a conformational search using a Mixed torsional/Large-Scale Low-Mode sampling method as implemented in MacroModel \citep{mohamadi_macromodelintegrated_1990}. For each identified conformer, the geometries were optimized, and zero-point corrected energies were calculated using Gaussian 16 \citep{g16} with the B3LYP/6-311++G(3df,2pd) functional and basis set. 
\cite{Masato2001} reported four possible forms of 3-hydroxypropanal, however, we obtained 13 conformations in the energy window of 16 kJ/mol. For the four most stable conformers, we then re-calculated the relative energies with the CCSD/aug-cc-pVDZ level of theory. The relative energies and dipole moments of these four conformers are displayed in Figure~\ref{fig:conformers}. 

Conformer 1 was predicted as the most stable form of 3-hydroxypropanal also by \cite{Masato2001}. It adopts a pseudo cyclic structure in which 
the hydroxyl hydrogen and carbonyl oxygen are in close proximity and thus allow for a stabilizing intramolecular hydrogen bonding interaction. 
This intramolecular hydrogen bonding results in Conformer 1 being 6.7\,kJ/mol (803 K) more stable than the next 
conformation which has an extended structure. Therefore, it is the only conformer that is expected to have a notable population in our spectrum. Of the four displayed conformers, Conformer 1, 3, and 4 all have two equivalent configurations connected by rotations around single bonds. This effective doubling increases the statistical weight of Conformer 1, further making it the predominant species in the gas phase of our sample.  To guide its assignment, rotational constants, centrifugal distortion constants, and electric dipole moment components were calculated at three different levels of theory: B2PLYPD3/6-311++G(3df,2pd) and MP2/aug-cc-pVTZ as implemented in the Gaussian 16 software \citep{g16}, and CCSD/aug-cc-pVDZ available in the CFOUR program package \citep{cfour}. The harmonic and anharmonic force field calculations were carried out in all cases. The convergence criteria for energies were set to 10$^{-11}$ atomic units. The obtained sets of spectroscopic parameters are provided in Table \ref{table:constants}. 
   \begin{figure*}[t]
   \centering
   \includegraphics[trim = 25mm 35mm 100mm 10mm, clip, width=16.0cm]{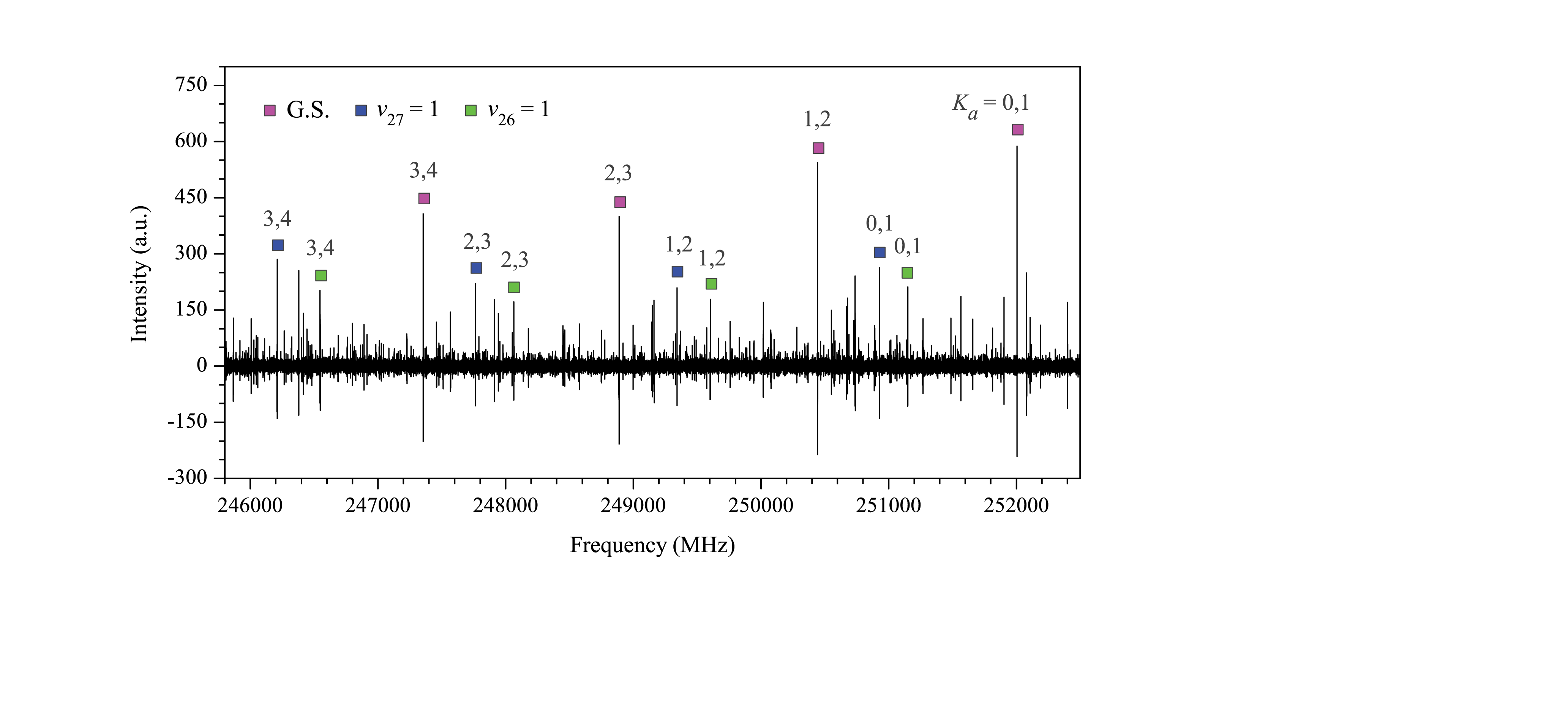}
      \caption{Section of the room-temperature millimeter wave spectrum of 3-hydroxypropanal. Dominating lines correspond to degenerate pairs of the ground state (G.S.) $a$-type and $b$-type R-branch transitions with low $K_a$ values of Conformer 1. The leading transition is $J = 40 \leftarrow 39$ and the value of $J$ decreases by 1 with each successive line. The same pattern is observable for the $v_{27}=1$ and $v_{26}=1$ excited vibrational states.}
      \label{fig:spectrum}
   \end{figure*}

\label{sec:calc}

\section{Experimental Description}
\label{sec:exp_description}
\subsection{Synthesis of 3-Hydroxypropanal}

The synthesis of \citet{roldan} was modified to obtain the product in the gas phase. 3,3-Diethoxy-1-propanol (1.0 g, 6.75 mmol) was slowly added to a cooled (0°C) 1 M aqueous solution of \ce{H2SO4} (5 mL). The reaction was stirred at 20 °C for 30 minutes. This mixture was then added to a suspension of \ce{NaHCO3} (6.3 g, 75 mmol) in dichloromethane (50 mL). At the end of bubbling, the organic phase was separated and the residual organic compounds in the aqueous phase extracted twice with dichloromethane (25 mL). The organic phases were mixed and dried over \ce{MgSO4}. After filtration, the solvent was removed at room temperature under vacuum until approximately 2-3 mL of solution was obtained. 3-hydroxypropanal was purified by trap-to-trap distillation under vacuum (0.1 mbar) and selective trapping in a U-tube immersed in a bath cooled at --40°C. (Yield: 0.17 g, 2.3 mmol, 34\%). Revaporization in the cell of the spectrometer enabled the millimeter wave spectrum to be recorded.

3-Hydroxypropanal is not very kinetically stable in pure form at room temperature in the condensed phase. The use of a crude mixture (obtained after evaporation of the solvent at 0°C under 0.1 mbar before connecting the sample to the spectrometer cell) stabilizes the product considerably, but ethanol was also observed in the spectrum. The yield of the product (based on the starting material 3,3-diethoxy-1-propanol) vaporized in the gas phase is no greater than 20\%.

The following is the nuclear magnetic resonance (NMR) data for the synthesized molecule:
$^{1}$H NMR (\ce{CDCl3}, 400 MHz) $\delta$ 9.87 (t, 1H, $^{3}$J\textsubscript{HH} = 1.0 Hz, CHO), 3.96 (t, 2H, $^{3}$J\textsubscript{HH} = 5.4 Hz, \ce{CH2O}), 2.77 (2H, td, $^{3}$J\textsubscript{HH} = 5.4, 1.0 Hz, \ce{CH2}). $^{13}$C NMR (\ce{CDCl3}, 100 MHz) $\delta$ 202.4 (CHO), 56.5 (\ce{CH2O}), 46.1 (\ce{CH2}).
\label{sec:exp_synthesis}

\subsection{Spectroscopic Experiments}
\label{sec:spectroscopy}

The rotational spectrum of 3-hydroxypropanal was measured in the frequency region 130--485~GHz using two spectrometers. The frequency-modulated absorption spectrum was first measured at room temperature over most of the frequency range spanning $150-190\,$GHz, $225-325\,$GHz, and $367-485\,$GHz using the FLASH  (Fast Lille Absorption emiSsion High resolution) spectrometer \citep{lille, fast_scan, zou2020}. Its fast scan design is achieved by mixing a direct digital synthesizer (DDS) signal with the signal from an Agilent synthesizer (up to 20 GHz), followed by a multiplication chain consisting of various passive and active multipliers. The FLASH spectrometer is particularly notable for incorporating the chirped-pulse technique. While this technique is well-established in the centimeter range, its application in the THz domain is still innovative, making the FLASH spectrometer a unique and versatile tool capable of operating up to 1.5 THz in absorption mode and 0.5 THz in chirped-pulse mode. Due to the instability of 3-hydroxypropanal, the measurements were conducted in ``flow" mode. In this configuration, the spectrometer features a 220 cm long Pyrex glass tube with a 10 cm diameter and Teflon windows. The sample was continuously injected into the cell and pumped out from the opposite side.


In this specific experiment, the frequency was modulated at $40.5\,$kHz, and $2f$ frequency demodulation was employed. Once again, because of the low molecular stability, the frequency range needed to be scanned fairly quickly. Thus, a frequency step 20\% larger than usual was used, ranging from 36 kHz at 150 GHz to 63 kHz at 400 GHz and above. The step size was increased at higher frequencies due to the heightened Doppler broadening of the lines. Throughout most of the frequency range, the spectra were averaged four times and the scanning rate was 1 ms/point. The sample pressure and temperature during measurements were about 2-3~Pa and  room temperature. Absorption signals were detected by zero-bias Schottky diodes. 

The Prague semiconductor millimeter wave spectrometer \citep{Kania2006} was also used to record the spectra in selected sections between 130 and 175~GHz. The required frequencies were obtained by multiplying the Agilent synthesizer produced frequency ($\leq$50 GHz) using a tripler or a pair of doublers. The synthesizer output was modulated at a frequency of 28~kHz. The stability of the sample signal was ensured by establishing a slow flow through the Pyrex free-space cell (8~cm in diameter, 280~cm long) at a pressure of around 4~Pa. The optical path length was doubled using a roof-top mirror. The signal was detected by zero-bias detectors and further processed by a lock-in amplifier that was tuned to twice the modulation frequency. The final spectrum was obtained as an average of scans sweeping forwards and backwards in a double acquisition cycle. 

The estimated frequency  uncertainties over the entire measurement range were either 30 kHz, 50 kHz, or, in a small number of cases, 100 kHz. These uncertainties were primarily determined by the signal-to-noise ratio of each line. For instance, strong b-type R branch transitions of the ground state generally had uncertainties of 30 kHz, while some very weak a-type lines were assigned uncertainties of 100 kHz.

\section{Spectroscopic Analysis and Results}
\label{sec:exp_results}

\subsection{Ground State}
The fit was initiated using a simulated spectrum that was generated with the theoretical ground state rotational constants and distortion constants using SPCAT in Pickett’s CALPGM suite of programs \citep{spcat}. Overall, the fit contains 1194 distinct transition frequencies. Including blended lines, this corresponds to 2342 transitions in the fit. The transitions range from $J^{\prime} = 8-76$ and $K_a^{\prime} = 0-40$. Because this molecule is a relatively prolate asymmetric top ($\kappa = -0.642$), the fit was conducted with the A reduction and $I^r$ representation using ASFIT \citep{asfit}. The values of each of the determined rotational and centrifugal distortion constants are listed in Table~\ref{table:constants}. Because Conformer 1 has a large b-type electric dipole moment component, 1794 of the fitted transitions are b-type. As displayed in Figure~\ref{fig:spectrum}, the strongest lines correspond to the low $K_a$ R-branch b-type transitions. Several of the high $K_a$ (e.g. $K_a{\prime} = 24-40$) Q-branch b-type transitions were also measured and fitted, albeit with much lower line intensities. The remaining 548 transitions in the fit were a-type lines; however, a fairly notable number of these are blended with b-type transitions (i.e. overlapping a-type and b-type R-branch transitions). The unblended a-type lines were generally very weak due to the smaller a-type dipole moment component. No c-type transitions were observed. The calculated rotational constants of Conformer 1 agree with the experimentally determined values within 3\% for each of the functional/basis set combinations. The signs and magnitudes of the computed centrifugal distortion constants also agree well with the experimental values. Following the fitting using ASFIT, the determined parameters were outputted by the program in the file format required for SPFIT/SPCAT. Thus, we subsequently used these resulting files to produce a spectroscopic catalog in the standard .cat format using SPCAT to compare the fitted spectrum (including line intensities) to both our experimental and astronomical data. The spectroscopic catalog along with both the ASFIT and SPCAT files are provided in the Supplementary Information. The predicted spectra are also available in different formats including standard .cat format \citep{spcat} from the new database of the Lille spectroscopy group called the Lille Spectroscopic Database\footnote{https://lsd.univ-lille.fr}. The predictions can be generated using various options (e.g., intensity units, temperature, and frequency range) that provide additional flexibility in the data access. The partition function of 3-hydroxypropanal is listed in Table~\ref{table:partition}. The rotational partition function was calculated with SPCAT by simulating the spectrum up to $J$ = 200. The vibrational partition function was computed using the approximation formalism presented in \citet{gordy_cooke} with the anharmonic CCSD/aug-cc-pVDZ fundamental vibrational mode frequencies. The total partition function was calculated as $Q_{\mbox{vib}} (T) \times Q_{\mbox{rot}} (T)$ at each temperature. 

Note that in the calculation of higher-order spectroscopic constants (e.g., centrifugal distortion and Coriolis coupling constants), the choice of quantum chemical method plays a critical role. Although B3LYP/6-311++G(3df,2pd) offers a clear advantage in terms of computational efficiency, it often falls short in accurately describing these properties due to its limited treatment of electron correlation. In a previous study on the FSO$_3$ radical \citep{Uhlikova_fso3_2011}, several computational approaches were evaluated for the calculation of higher-order spectroscopic constants. The results demonstrated that CCSD yielded the best agreement with experimental data, further highlighting the limitations of B3LYP in this context and reinforcing the theoretical robustness of coupled-cluster methods for spectroscopic applications.
Nevertheless, using the anharmonic fundamental vibrational frequencies from the other two methods in Table~\ref{table:constants} results in vibrational partition functions that differ from CCSD by less than 2\% at 300~K.

\begin{table*}
\centering
\caption{Experimental spectroscopic parameters of Conformer 1 of 3-hydroxypropanal along with theoretical counterparts.}
\label{table:constants}
\begin{threeparttable}
\setlength{\tabcolsep}{9pt}
\begin{tabular}{llllllll}
\hline
 & \multicolumn{3}{c}{Experiment} & & \multicolumn{3}{c}{Theory\tnote{a}} \\
\cline{2-4} 
\cline{6-8}
Parameter & Ground State & $v_{27}=1$ & $v_{26}=1$ & & I & II & III \\
\hline
$A$ (MHz) & $8657.60135(34)$        & 8693.8551(12) & 8688.5405(12)& &   8513  &    8661   &    8606  \\ 
$B$ (MHz) &  $4116.27295(17)$        & 4083.01(26)& 4087.07(26)& &   3993  &    4054   &    4103  \\ 
$C$ (MHz) & $3125.63135(11)$        & 3112.7252(10) & 3114.7592(12)& &   3049  &    3089   &    3118  \\ 
\\                                                                          
$\Delta_{J}$ (kHz) & $4.010060(69)$ & 4.4948(78)& 3.6876(77)& &   3.663 &   3.675   &   3.834  \\ 
$\Delta_{JK}$ (kHz) &$-9.53279(46)$& $-$11.687(27)& $-$8.722(28)& & $-$8.238 &$-$8.280   & $-$8.344  \\ 
$\Delta_{K}$ (kHz) & 12.99819(71)  & 16.299(20)& 12.487(22) & &  11.168 &  11.535   &  11.033  \\ 
$\delta_{J}$ (kHz) &1.274532(30) & 1.4779(40)& 1.1381(39)& &   1.142 &   1.156   &   1.200  \\ 
$\delta_{K}$ (kHz) & 5.30738(43)  & [5.30738]\tnote{c}& 5.7674(25)& &   4.851 &   4.966   &   4.937  \\ 
\\                                                                          
$\Phi_{J}$ (Hz) &$-$0.019624(13)   & $-$0.01875(10)& $-$0.019507(11)& & $-$0.013 & $-$0.010   & $-$0.016  \\ 
$\Phi_{JK}$ (Hz) & 0.12908(25)    & 0.10550(85)&  0.16475(63)& &   0.095 &   0.086   &   0.110  \\ 
$\Phi_{KJ}$ (Hz) &$-$0.28906(53)   & $-$0.3020(18)& $-$0.3456(19)& & $-$0.223 & $-$0.220   & $-$0.241  \\ 
$\Phi_{K}$ (Hz) & 0.25706(76)     & 0.3476(17)& [0.25706]& &   0.194 &   0.200   &   0.199  \\ 
$\phi_{J}$ (Hz) &$-$0.0091793(59)  & $-$0.008748(50)& [$-$0.0091793]& & $-$0.006 & $-$0.005   & $-$0.007  \\ 
$\phi_{JK}$ (Hz) &$-$0.06190(11)   & $-$0.07923(54)& $-$0.05572(66)& & $-$0.037 & $-$0.032   & $-$0.041  \\ 
$\phi_{K}$ (Hz) &$-$0.08147(63)    & [$-$0.08147] & [$-$0.08147]& & $-$0.037 & $-$0.018   & $-$0.046  \\ 
\\                                                                       
$L_{JJK}$ (mHz) & 0.001124(38)    & [0.001124]& [0.001124]& &  ...    &   ...     &   ...    \\ 
$L_{JK}$ (mHz) &$-$0.00480(11)     & [$-$0.00480]& [$-$0.00480]& &  ...    &   ...     &   ...    \\ 
$L_{KKJ}$ (mHz) & 0.00828(15)     & [0.00828]& [0.00828]& &  ...    &   ...     &   ...    \\ 
$L_{K}$ (mHz) & $-$0.00664(25)     & [$-$0.00664]& [$-$0.00664]& &  ...    &   ...     &   ...    \\ 
\\

$G_a$ (MHz) & ...& \multicolumn{2}{c}{[1559.2]\tnote{d}} & & 1559.2\tnote{e} & 1544.6 & 1510.2 \\ 
$G_b$ (MHz) &... & \multicolumn{2}{c}{3088(54)} & & 2433.1\tnote{e} & 2463.8 & 2465.3 \\ 
$G_b^J$ (MHz)  & ...& \multicolumn{2}{c}{$-$0.0534(23)} & & ... & ... & ... \\ 
$G_b^K$ (MHz)  & ...& \multicolumn{2}{c}{0.1065(28)} & & ... & ... & ... \\ 
$\Delta E$ (cm$^{-1}$)\tnote{f}  & ... & \multicolumn{2}{c}{43.8611(18)} &  & 43.5 & 69.3 & 64.1 \\ 
\\
$|\mu_a|$ (D) & ... & ... & ...  & &  1.0    &   1.0     &   1.0    \\ 
$|\mu_b|$ (D) & ... & ... & ...  & &  2.8    &   2.8     &   2.8    \\ 
$|\mu_c|$ (D) & ... & ... & ...  & &  0.1    &   0.1     &   0.2    \\ 
\\
$N\textsubscript{lines}$\tnote{b}  & 1194  & 484 & 316 & &  ... & ... &   ...    \\ 
$\sigma\textsubscript{rms}$ (MHz)  & 0.0420 & \multicolumn{2}{c}{0.0635}  & & &  ... & ...   \\ 

$\sigma\textsubscript{rms,weighted}$  & 0.84 & \multicolumn{2}{c}{0.97}  & & &  ... & ...   \\ 
\hline
\end{tabular}
    \begin{tablenotes}
\item[a] I: CCSD/aug-cc-pVDZ, II: B2PLYPD3/6-311++g(3df,2pd), III: MP2/aug-cc-pVTZ. Rotational and centrifugal distortion constants correspond to the ground state.
\item[b] Number of distinct frequencies included in the fit.
\item[c] Bracketed distortion constants were fixed to the ground state values.
\item[d] Fixed to value calculated at CCSD/aug-cc-pVDZ functional and basis set.
\item[e] The $G_a$ and $G_b$ parameters from theory were determined using Equation~\ref{eq:coriolis} with the computed $\zeta$ and energy values and the experimental ground state rotational constants.
\item[f] Energy difference between the $v_{27}=1$ and $v_{26}=1$ states.
\end{tablenotes}
\end{threeparttable}
\end{table*}

\begin{table}
\centering
\caption{Partition function of Conformer 1 of 3-hydroxypropanal.}
\label{table:partition}
\begin{tabular}{cccc}
\hline
Temperature (K) & $Q_{\mbox{rot}} (T)$ & $Q_{\mbox{vib}} (T)$ & $Q_{\mbox{tot}} (T)$ \\
\hline
300.000 & 83321.061 & 8.99 & 749274.02\\
225.000 & 54070.707 & 4.20 & 227222.08\\
150.000 & 29408.229 & 2.06 & 60609.93\\
75.000 & 10391.237 & 1.16 & 12017.88\\
37.500 & 3674.599 & 1.01 & 3711.28\\
18.750 & 1300.583 & 1.00 & 1300.67\\
9.375 & 460.987 & 1.00 & 460.99\\

\hline
\end{tabular}
\end{table}

\subsection{Vibrationally Excited States}

The two lowest energy vibrationally excited states were also clearly visible in the spectrum. These were the $v_{27}=1$ and $v_{26}=1$ states. The corresponding normal vibrational modes are fairly similar torsional vibrations. The $v_{27}$ mode is predominantely a torsion around the C--C bond containing the aldehyde carbon. The $v_{26}$ mode is then a torsional motion around both C--C bonds with differing proportions. The two states were calculated to be 125.0 and 168.5~cm$^{-1}$ above the ground state, respectively. This computed difference of only 43.5 cm$^{-1}$ (63K, 0.52~kJ/mol) resulted in some observable Coriolis coupling between the two states that impacted a fairly small number of transitions. Therefore, these states were fitted with a combined fit using SPFIT code \citep{spcat}. The PIFORM code \citep{kisiel_assignment_2001} was employed to format the output.

\begin{figure}[h!]

\begin{center}
\includegraphics[width=\columnwidth]{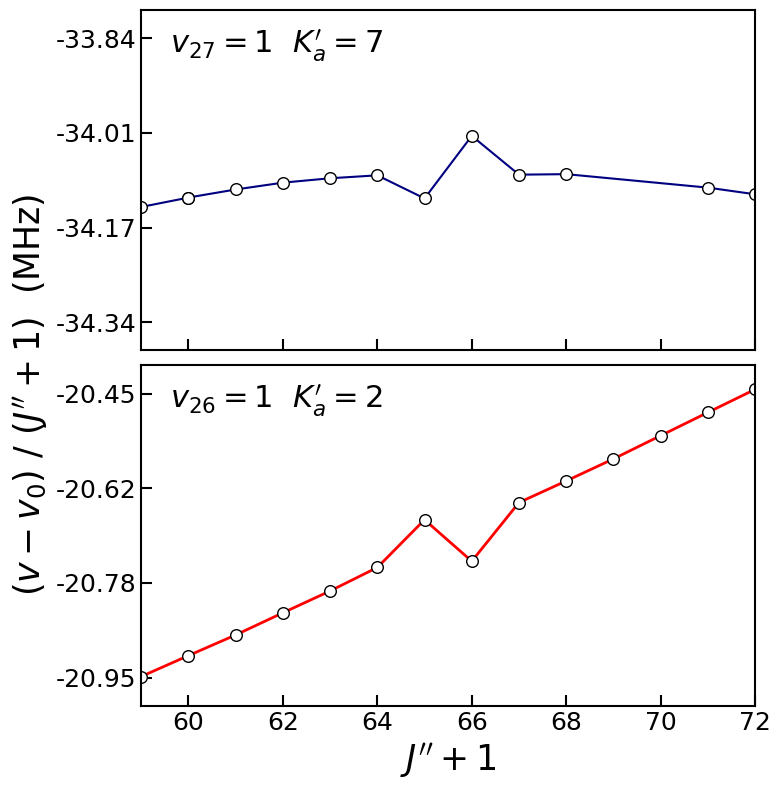}
\caption{Depiction of some perturbed transitions in the 
$v_{27}=1$ and $v_{26}=1$ states. The y axis depicts the difference between the transition frequency in the excited and ground vibrational state, scaled by $J^{\prime\prime} + 1$. All transitions are b-type R branch lines. The scatter points depict the measured transitions. The red and blue lines connect each of the scatter points in order to more clearly display the perturbation. For the values without points, either the ground state or excited state transition occurred at a frequency that was not measured in our experiment. }
\label{fig:perturbation}
\end{center}
\end{figure}

In general, the lines that displayed a notable and observable perturbation were typically the low-to-mid $K_a$ R-branch transitions. The most significant perturbations were Coriolis couplings of $\Delta K_a = 5$ beginning at $K_a^{\prime} = 1$ for $v_{26}=1$ and $K_a^{\prime} = 6$ for $v_{27}=1$. The frequency perturbations of the transition frequencies generally ranged from approximately 300 kHz to 4 MHz. An example of these perturbed transitions is displayed in the resonance plot in Figure~\ref{fig:perturbation}. Unfortunately, however, while there were clear perturbations of $v_{26}=1$ $K_a^{\prime} =$ 5 and 6 transitions, the perturbing counterparts $K_a^{\prime} = $ 10, 11, and 12 lines of $v_{27}=1$ became too weak to observe. Thus, there were only a very small number of paired perturbing transitions that we were able to include in the fit, making it challenging to constrain the required Coriolis coupling parameters. Additionally, since the largest observed perturbations in the transition frequencies were slightly above 4 MHz, these fairly small deviations further complicated the task of constraining the coupling parameters.

The fit was initialized with the computed energy difference between the two states, and the $G_a$ and $G_b$ parameters calculated using the following general equation:

\begin{equation}
G_a = \frac{\omega_{27}+\omega_{26}}{\sqrt{\omega_{27}\times\omega_{26}}}\zeta^{a}_{27,26}A
\label{eq:coriolis}
\end{equation}
where $\omega$ is the computed vibrational frequency of the two interacting modes, and $\zeta$ was computed from the anharmonic CCSD/aug-cc-pVDZ calculations. For computing $G_b$, Equation~\ref{eq:coriolis} is used with $\zeta^{b}_{26,27}$ and $B$ substituted in. The fitting procedure was carried out using a conventional bootstrap approach. Initially, the parameter $\Delta E$ was fixed to a range of values between 35 and 55~cm$^{-1}$, while the parameter $G_b$ was allowed to vary. We then searched for a global minimum of the RMS deviation, following the method described by \citet{motiyenko_internal_2018}. Once the global minimum was identified, $\Delta E$ was also treated as an adjustable parameter. Given the limited number of observed perturbed transitions, the final fit was performed by fixing $G_a$ to its computed value while allowing $G_b$, $\Delta E$, $G_b^{J}$, and $G_b^{K}$ to vary. $G_b^{J}$ and $G_b^{K}$ are centrifugal distortion correction terms to the $G_b$ Coriolis coupling constant. The frequency uncertainty of the perturbed transitions was extended to 100~kHz in the fit.

Ultimately, a total of 2172 transitions were included in the excited-state fit, corresponding to 800 distinct lines. The fitted rotational, distortion, and Coriolis coupling constants are displayed in Table~\ref{table:constants}. The perturbed transitions are much more accurately reproduced using the two-state fit with Coriolis coupling parameters rather than single-state fits. However, some of these lines still exhibit notable deviations from the fit, leading to a relatively large RMS deviation of approximately 63 kHz. The unblended a-type transitions, which were already extremely weak in the ground state, are too faint to be observed in the spectrum for the vibrationally excited states. The energy difference between the two states from the fit is exceptionally close to the difference of the computed anharmonic fundamental frequencies from the CCSD/aug-cc-pVDZ calculation. Each of the fitted centrifugal distortion constants in the excited states have the same sign and similar magnitude to the ground state values. However, as can be seen, several of the centrifugal distortion constants in the two vibrationally excited states have fairly equal and opposite differences from the ground-state value. This is likely suggesting that a notable amount of the Coriolis coupling is being absorbed into the centrifugal distortion constants. If greater signal-to-noise ratio could be achieved in future work, more perturbed transitions could be included in the fit and the determined distortion and Coriolis-coupling constants could therefore be improved. However, due to the very unstable nature of the sample, this would likely be a challenging task. 

\section{Observational Searches and Results}
\label{sec:obs_results}

\subsection{G+0.693-0.027 Molecular Cloud}
\label{g693_description}
We have searched for 3-hydroxypropanal towards the Galactic Center molecular cloud G+0.693-0.027 (hereafter G+0.693), which is located in the Sgr B2 complex. 
The molecular clouds in the Galactic Centre exhibit high gas kinetic temperatures ranging from $\sim$50 to $\sim$150 K (\citealt{guesten1985,huettemeister1993,rodriguez-fernandez2001,ginsburg2016,krieger2017}), low dust temperatures of $\leq$ 20$-$30 K (\citealt{rodriguez-fernandez2004,battersby2024}), and relatively low H$_2$
gas densities around $10^{4}\text{--}10^{5}$ cm$^{-3}$ (\citealt{guesten1983,bally1987b,rodriguez-fernandez2000}). 
The G+0.693 cloud, located around 55$^{\prime\prime}$ northeast of Sgr B2(N), is not an exception, and exhibits high kinetic temperatures of 70$-$150 K (\citealt{Zeng2018}), and densities of 10$^{4\text{--}5}$ cm$^{-3}$ (\citealt{zeng2020,colzi2022,colzi2024}). 

\begin{center}
\begin{figure*}[ht]
     \centerline{\resizebox{0.90
     \hsize}{!}{\includegraphics[angle=0]{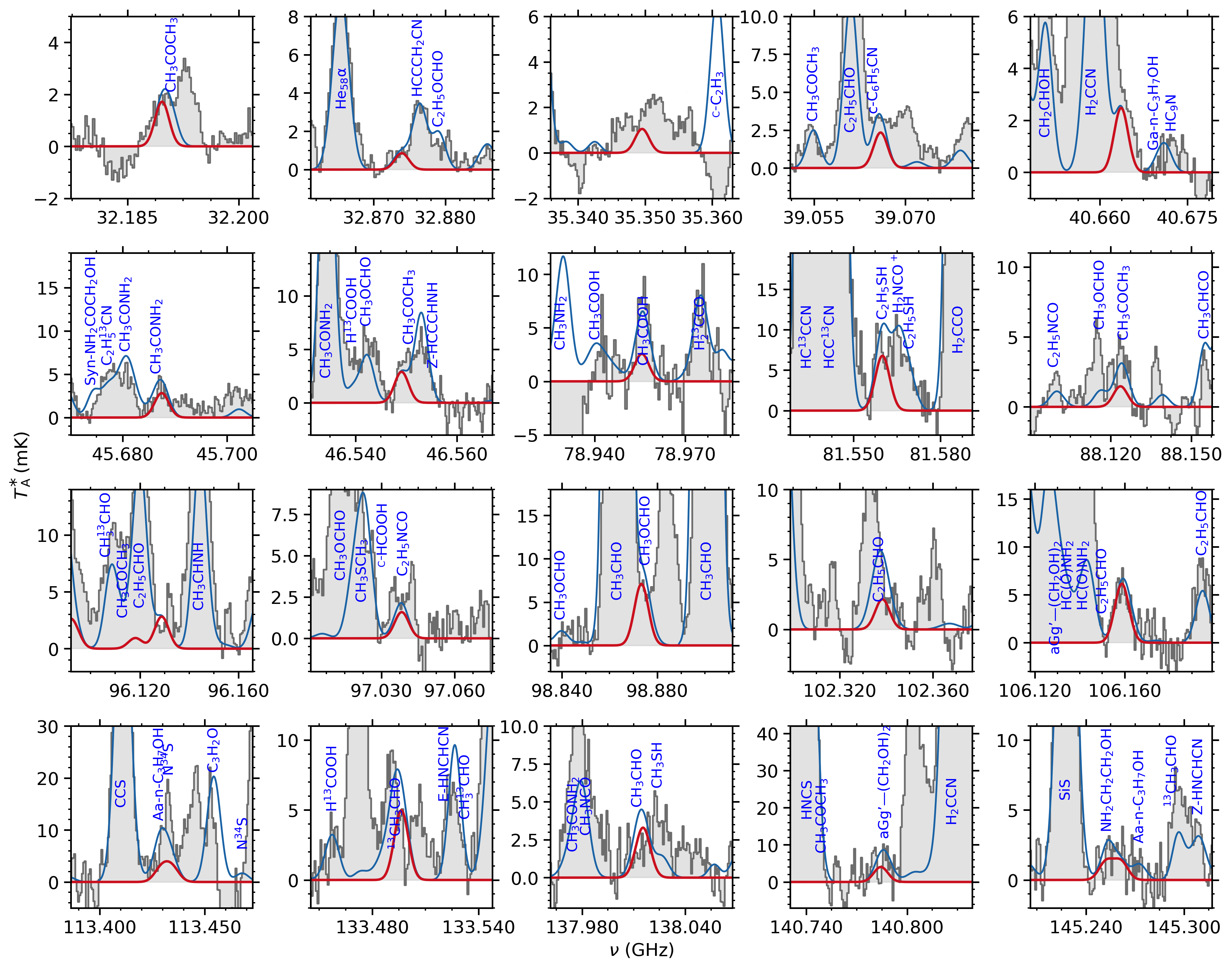}}}
     \caption{Tentative detection of 3-hydroxypropanal toward G+0.693. The selected transitions are sorted by increasing frequency and listed in Table \ref{tab:transitions693}. The result of the LTE model of 3-hydroxypropanal is plotted with a red line  while the blue line represents the combined emission of all molecules identified in the survey, including also 3-hydroxypropanal (observed spectra shown as gray lines and light gray-shaded histograms).}
\label{f:LTEspectrumG0693}
\end{figure*}
\end{center}

\begin{table*}
\centering
\tabcolsep 3pt
\caption{Spectroscopic information of the transitions of 3-hydroxypropanal tentatively detected toward G+0.693$-$0.027 (shown in Figure \ref{f:LTEspectrumG0693}).}
\begin{tabular}{ccccccccccc}
\hline\hline
Frequency & Transition $^{(a)}$ & log \textit{I}& \textit{g}$\mathrm{_u}$ & $E$$\mathrm{_{up}}$ &  Blending  \\ 
(GHz) & &  (nm$^2$ MHz) &  &  (K)  & \\
\hline
32.1896471 (13)  & 5$_{0,5}$ -- 4$_{1,4}$  & --5.6326  & 11  & 5.0 &  Slightly blended: \ch{CH3COCH3} \\					
32.8739195 (29)  & 7$_{4,3}$ -- 7$_{3,4}$  & --5.6233  & 15  &13.6 &  Slightly blended: \ch{HCCCH2CN} \\ 					
35.3495481 (13)  & 3$_{2,2}$ -- 2$_{1,1}$  & --5.8750  &  7  & 3.0 &  U-line \\ 
39.0659430 (15)  & 6$_{0,6}$ -- 5$_{1,5}$  & --5.3517  & 13  & 6.9 &  $c$-\ch{C6H5CN} \\				
40.6636428 (15)  & 6$_{1,6}$ -- 5$_{0,5}$  & --5.3091  & 13  & 6.9 &  \ch{H2CCN} \\ 				
45.6874831 (17)  & 7$_{0,7}$ -- 6$_{1,6}$  & --5.1292  & 15  & 9.1 &  Slightly blended: \ch{CH3CONH2} \\  						
46.5490784 (17)  & 7$_{1,7}$ -- 6$_{0,6}$  & --5.1103  & 15  & 9.1 &  Unblended \\  						
78.9556730 (29)  & 6$_{4,2}$ -- 5$_{3,3}$  & --4.8794  & 13  & 11.1 &  \ch{CH3COOH} \\ 		
81.5594297 (34)  & 5$_{5,1}$ -- 4$_{4,0}$  & --4.7334  & 11  & 11.2 &  \ch{C2H5SH} \\ 
81.5605329 (34)  & 5$_{5,0}$ -- 4$_{4,1}$  & --4.7334  & 11  & 11.2 &  \ch{C2H5SH} \\ 		
88.1236563 (27)  &13$_{1,12}$ -- 12$_{2,11}$ & --4.3761  & 27  & 31.2 &  \ch{CH3COCH3} \\ 		
96.1288496 (35)  & 7$_{5,2}$ -- 6$_{4,3}$  & --4.6019  & 15  & 15.7 &  Unblended \\		
97.0383121 (32)  & 9$_{4,6}$ -- 8$_{3,5}$  & --4.7138  & 19  & 19.5 & Slightly blended: \ch{C2H5NCO} \\ 	
98.8733057 (40)  & 6$_{6,1}$ -- 5$_{5,0}$  & --4.4842  & 13  & 11.2 &  \ch{CH3OCHO} \\ 		
98.8733818 (40)  & 6$_{6,0}$ -- 5$_{5,1}$  & --4.4842  & 13  & 11.2 &  \ch{CH3OCHO} \\ 		
102.3378081 (33) &16$_{0,16}$ -- 15$_{1,15}$  & --4.0589  & 33  & 42.3 &  \ch{s-C2H5CHO} \\ 
102.3381298 (33) &16$_{1,16}$ -- 15$_{1,15}$  & --4.9124  & 33  & 42.3 &  \ch{s-C2H5CHO} \\ 			
102.3385252 (33) &16$_{0,16}$ -- 15$_{0,15}$  & --4.9124  & 33  & 42.3 &  \ch{s-C2H5CHO} \\ 
102.3388468 (33) &16$_{1,16}$ -- 15$_{0,15}$  & --4.0589  & 33  & 42.3 &  \ch{s-C2H5CHO} \\
106.1583426 (40) & 7$_{6,2}$ -- 6$_{5,1}$  & --4.4294  & 15  & 18.3 &  Unblended \\ 	
106.1591837 (40) & 7$_{6,1}$ -- 6$_{5,2}$  & --4.4294  & 15  & 18.3 &  Unblended \\ 		
113.4294989 (40) & 8$_{6,3}$ -- 7$_{5,2}$ & --4.3771  & 17  & 21.1 &  \ch{N$^{34}$S} and $n$-\ch{C3H7OH} \\ 	
113.4345521 (40) & 8$_{6,2}$ -- 7$_{5,3}$ & --4.3771  & 17  & 21.1 &  \ch{N$^{34}$S} and $n$-\ch{C3H7OH} \\ 
133.4964353 (49) & 8$_{8,1}$ -- 7$_{7,0}$ & --4.1035  & 17  & 27.8 &   \ch{$^{13}$CH3CHO} \\ 
133.4964356 (49) & 8$_{8,0}$ -- 7$_{7,1}$ & --4.1035  & 17  & 27.8 &   \ch{$^{13}$CH3CHO}\\ 
138.0143053 (45) &10$_{7,4}$ -- 9$_{6,3}$ & --4.1474  & 21  & 30.8 &  \ch{CH3CHO} \\ 		
138.0165330 (45) &10$_{7,3}$ -- 9$_{6,4}$ & --4.1474  & 21  & 30.8 &  \ch{CH3CHO} \\ 
140.7843737 (48) & 9$_{8,2}$ -- 8$_{7,1}$ & --4.0650  & 19  & 30.9 &  AgG'-(CH$_2$OH)$_2$ \\ 
140.7843783 (48) & 9$_{8,1}$ -- 8$_{7,2}$ & --4.0650  & 19  & 30.9 &  AgG'-(CH$_2$OH)$_2$ \\		
145.2523185 (44) &11$_{7,5}$ -- 10$_{6,4}$ & --4.1089  & 23  & 34.7 &  \ch{NH2CH2CH2OH} \\ 		
145.2612631 (45) &11$_{7,4}$ -- 10$_{6,5}$ & --4.1088  & 23  & 34.7 &  \ch{NH2CH2CH2OH} \\ 	
\hline 
\end{tabular}
\label{tab:transitions693}
\tablecomments{$^{(a)}$ We used the conventional notation for asymmetric tops: $J_{K_{a},K_{c}}$ to label the rotational energy levels, where $J$ denotes the angular momentum quantum number, and the $K_{a}$ and $K_{c}$ labels are projections of $J$ along the $a$ and $c$ principal axes. Numbers in parentheses represent the predicted uncertainty associated to the last digits.}
\end{table*}

Due to the low H$_2$ densities, the molecules are generally sub-thermally excited, and hence the excitation temperatures in most cases are significantly lower than the gas kinetic temperatures ($\sim$5$–$20 K; \citealt{requena-torres2006,martin2008tracing,Zeng2018,Jimenez-Serra2020,Rivilla2022_nitriles}). For this reason, the resulting spectra are less crowded than in high-$T_{\rm ex}$ sources such as hot cores/corinos, which alleviates line confusion, and helps the identification of new interstellar molecules.  
Moreover, the region harbors large-scale shocks, as seen through the bright emission of SiO and HNCO, which has been attributed to a cloud-cloud collision (\citealt{zeng2020}). These shocks have sputtered the icy mantles of dust grains, ejecting into the gas phase  molecules formed via grain surface chemistry. Furthermore, the high cosmic-ray ionization rate of the cloud, 10$^{-14}-$10$^{-15}$  s$^{-1}$, inferred by the chemical modeling of molecular cations such as PO$^+$ and HOCS$^+$ (\citealt{Rivilla2022_PO+,Sanz-Novo2024a}), has been proposed to trigger the formation of reactive radicals on the ices, which can rapidly combine in-situ to synthetize more complex molecules (\citealt{rivilla_first_2023}). From all this, G+0.693 stands as one of the most chemically rich reservoirs in the Galaxy, and nearly 25 new interstellar species have been discovered towards this source (e.g. \citealt{Rivilla2020b,Rivilla2021a,rivilla_first_2023}), including complex O-bearing  species such as (\textit{Z})-1,2-ethenediol ((CHOH)$_2$, \citealt{Rivilla2022a}), propanol (C$_3$H$_7$OH; \citealt{Jimenez-Serra2022}), and carbonic acid (HOCOOH; \citealt{Sanz-Novo2023}). Moreover, other O-bearing molecules such as propenal or propanal are abundant toward this source (\citealt{requena-torres_galactic_2008}), which makes G+0.693 a good target for the interstellar search of 3-hydroxypropanal.

We used an ultra-sensitive spectral survey towards this cloud that comprised $Q$-band observations at centimeter wavelengths (31.075-50.424 GHz), conducted with the Yebes 40$\,$m radiotelescope (Guadalajara, Spain), and observations across several millimeter-wavelength frequency windows (83.2–115.41 GHz, 132.28–140.39 GHz, and 142.00–173.81 GHz), conducted with the 30$\,$m IRAM radio telescope in Granada (Spain). The position switching mode was used, centered at $\alpha$ = $\,$17$^{\rm h}$47$^{\rm m}$22$^{\rm s}$, $\delta$ = $\,-$28$^{\circ}$21$^{\prime}$27$^{\prime\prime}$, with the off position shifted by $\Delta\alpha$~=~$-885$$^{\prime\prime}$ and $\Delta\delta$~=~$290$$^{\prime\prime}$. 
The half power beam width (HPBW) of the Yebes 40$\,$m telescope varies between $\sim$35$^{\prime\prime}$-55$^{\prime\prime}$ (at 50 and 31 GHz, respectively; \citealt{Tercero2021}) and the HPBW of the IRAM 30$\,$m radiotelescope is $\sim$14$^{\prime\prime}$$-$29$^{\prime\prime}$ across the frequency range covered. The line intensity of the spectra is presented in units of antenna temperature ($T_{\mathrm{A}}^{\ast}$) since the molecular emission towards G+0.693 is extended beyond the telescope primary beams (\citealt{Brunken2010,Jones2012,Li2020,Santa-Maria2021,colzi2024}). Further details on these observations, including resolution and final noise levels of the molecular line survey, are given in \citet{rivilla_first_2023} and \citet{Sanz-Novo2023}. We note that for the frequency ranges not covered by these new observations, we used data from our previous IRAM 30m survey (see \citealt{Rodriguez-Almeida2021a} for additional information).


We implemented the spectroscopic entry of 3-hydroxypropanal into the analysis software \textsc{Madcuba}\footnote{Madrid Data Cube Analysis on ImageJ is a software developed at the Centre of Astrobiology (CAB) in Madrid: \url{https://cab.inta-csic.es/madcuba/}; see \citep{Martin2019}}. Using the Spectral Line Identification and Modelling (SLIM) tool of the \textsc{Madcuba} package \citep{Martin2019}, we generated the expected local thermodynamic equilibrium (LTE) synthetic spectra of 3-hydroxypropanal to search for it in the G+0.693 spectral survey. We note that, given the low excitation temperatures of the molecules in G+0.693 (as described above), we did not include the vibrational contribution to the partition function ($Q_{\mbox{vib}}$) in this case.

In Figure \ref{f:LTEspectrumG0693} we show the brightest transitions of 3-hydroxypropanal, according to the LTE modeled spectra, some of which have been tentatively identified toward G+0.693. Their spectroscopic information is reported in Table \ref{tab:transitions693}. Besides the clear detection of several relatively bright $b$-type $R$-branch lines, only a few appear to be unblended or only slightly blended with the emission from other molecules previously detected in this cloud. We highlight the 7$_{6,2}$ -- 6$_{5,1}$ and 7$_{6,1}$ -- 6$_{5,2}$ transitions (located at 106.1483426 GHz and 106.1591837 GHz, respectively) which coalesce into a single, stronger spectral feature owing to the broad linewidths of the molecular emission observed in this source (FWHM $\sim$ 15$-$20 km s$^{-1}$; \citealt{Requena-Torres2008,Zeng2018,Rivilla2022_nitriles}), and the 7$_{0,7}$ -- 6$_{1,6}$ transition (located at 45.6874831 GHz) that is only slightly contaminated by \ch{CH3CONH2}. Overall, the inclusion of 3-hydroxypropanal greatly enhances the agreement between the model and the observed spectra (see Figure \ref{f:LTEspectrumG0693}). Moreover, even with a non-negligible line blending, 3-hydroxypropanal appears to be the main carrier of most of the observed spectral features (e.g., the lines at $\sim$81.560 GHz and $\sim$ 39.066 GHz), further supporting its tentative detection. In Figures \ref{f:LTEall1} and \ref{f:LTEall2}, we present the remaining brightest transitions of 3-hydroxypropanal covered by the survey, not shown in Figure 4, corresponding to a total of $\sim$45 additional lines. As can be seen in both figures, the majority of these lines are severely contaminated by emission from other molecules previously identified in this survey, yet they do not show clear discrepancies between the model and the observed data.

To derive the column density, we employed the \textsc{Autofit}-SLIM tool \citep{Martin2019} and conducted a nonlinear least-squares LTE fit of the 3-hydroxypropanal emission to the observed spectra. We used the subset of transitions reported in Table \ref{tab:transitions693}, taking into account the contribution to the emission from other species. We adopted the excitation temperature ($T_{\rm ex}$), radial velocity ($v_{\rm LSR}$) and line width (FWHM) values derived for $s$-propanal in \citet{sanz-novo2022a}: $T_{\rm ex}$ = 12 K, $v_{\rm LSR}$ = 69 km s$^{-1}$ and FWHM = 21 km s$^{-1}$, leaving the column density as the only free parameter in the fit. We obtained a $N$ = (8.6$\pm$1.4)$\times$10$^{12}$ cm$^{-2}$, which translates into a molecular abundance with respect to H$_2$ of (6.4$\pm$1.4) $\times$10$^{-11}$, adopting a $N_{\rm H_2}$ = 1.35$\times$10$^{23}$ cm$^{-2}$ \citep{martin2008tracing} and assuming an uncertainty of 15\% of its value. The derived value not only accurately constrains the abundance of 3-hydroxypropanal in the ISM but also implies that this species is a factor of 9$\pm$2 less abundant than $s$-propanal in G+0.693 ($N$ = (7.4$\pm$1.5$\times$10$^{13}$ cm$^{-2}$) and a factor of $\sim$60 less abundant than acetaldehyde (5.0$\pm$0.1$\times$10$^{14}$ cm$^{-2}$) \citep{sanz-novo2022a}. Therefore, we observed a drop in abundance of nearly one order of magnitude as chemical complexity increased (i.e., with the addition of an OH group), which is in remarkable agreement with the results obtained for the related $n$-butanal toward the same molecular cloud ($s$-propanal/$ct$-$n$-butanal $>$ 9; \citealt{sanz-novo2022a}). The column densities of 3-hydroxypropanal, acetaldehyde and propanal in each of the sources investigated in this work are displayed in Table~\ref{table:column_densities}.

\subsection{Sgr~B2(N) Star-Forming Region}
\label{s:sgrb2}

Sagittarius (Sgr) B2(N) is a high-mass star forming protocluster located in 
the molecular cloud complex Sgr~B2 in the central molecular zone of our 
Galaxy, at a distance of 8.2~kpc \citep[][]{Reid19}. Sgr~B2(N) contains 
numerous hot molecular cores and H\,\textsc{ii} regions 
\citep[][]{Gaume95,Bonfand17,SanchezMonge17}. It was the target of the imaging 
spectral line survey Reexploring Molecular Complexity with ALMA (ReMoCA) that 
was performed with the Atacama Large Millimeter/submillimeter Array (ALMA) between 84.1 and 114.4~GHz. Details about the data 
reduction and the method of analysis can be found in 
\citet{Belloche19,2025A&A...698A.143B}. The survey was analyzed by comparing the 
observed spectra to synthetic spectra computed assuming LTE with the software Weeds \citep[][]{Maret11}. For this work
we focused on the position located at ($\alpha, \delta$)$_{\rm J2000}$= 
($17^{\rm h}47^{\rm m}19{\fs}87, -28^\circ22'19{\farcs}48$). This position, 
called Sgr~B2(N1S), is about 1$\arcsec$ to the south of the main hot core 
Sgr~B2(N1). It was chosen by \citet{Belloche19} as a compromise between 
reducing the dust optical depth and keeping a high enough H$_2$ column density 
to detect less abundant molecules.

\begin{figure*}
\centerline{\resizebox{0.85\hsize}{!}{\includegraphics[angle=0]{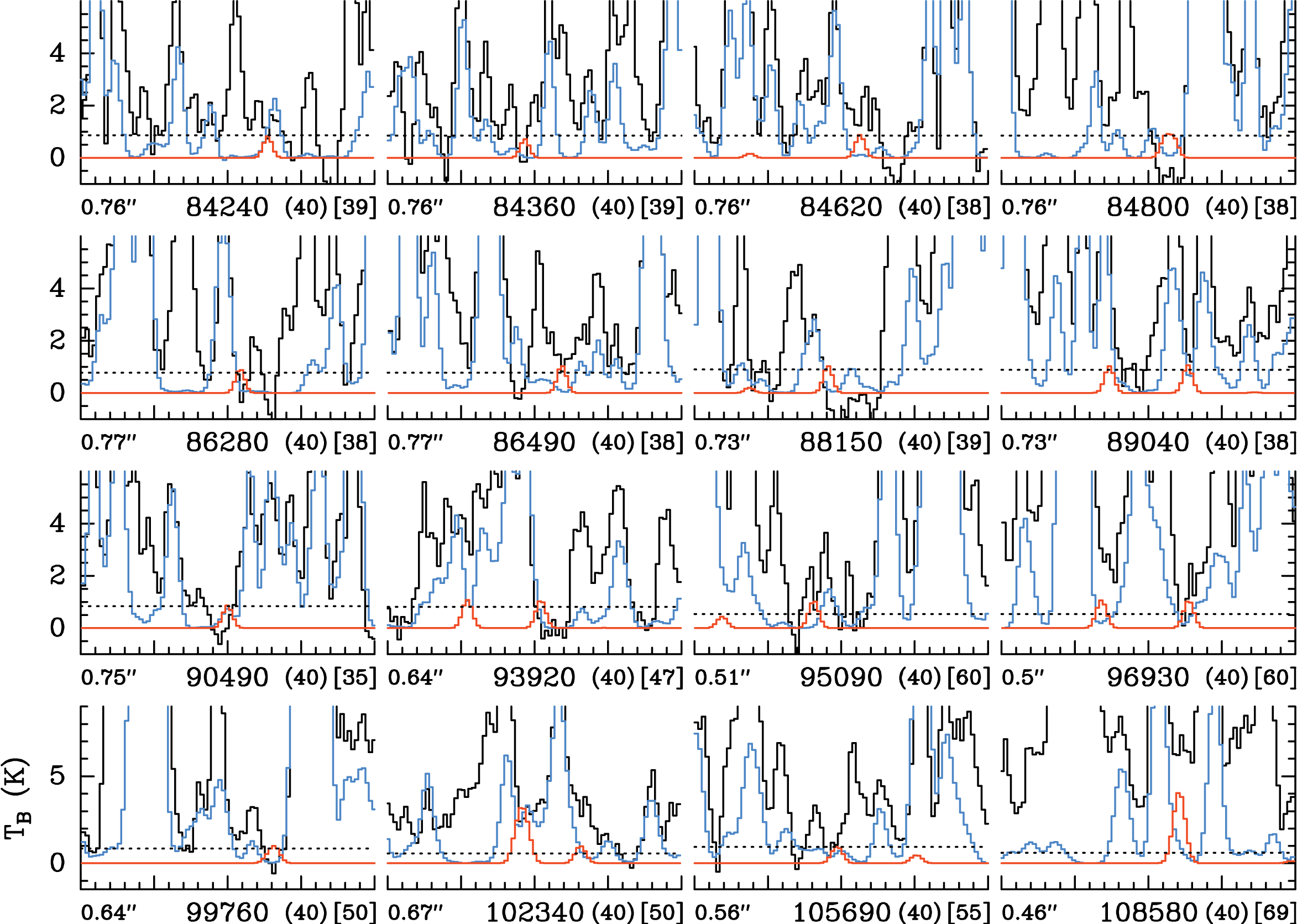}}}
\caption{
Selection of rotational transitions of 3-hydroxypropanal, $v$=0 covered by 
the ReMoCA survey. The LTE synthetic spectrum of 3-hydroxypropanal, $v$=0 
used to derive the upper limit to its column density toward 
Sgr~B2(N1S) is displayed in red and overlaid on the 
observed spectrum shown in black. The blue synthetic spectrum contains the 
contributions of all molecules identified in our survey so far. The 
values written below each panel correspond from left to right to the half-power 
beam width, the central frequency in MHz, the width in MHz of each panel in 
parentheses, and the continuum level in K of the baseline-subtracted spectra 
in brackets. The y-axis is labeled in brightness temperature units (K). The 
dotted line indicates the $3\sigma$ noise level.}
\label{f:remoca_hoch2ch2cho_ve0_n1s}
\end{figure*}

In order to search for 3-hydroxypropanal toward Sgr~B2(N1S), we assumed the 
same LTE parameters as acetaldehyde ($T_{ex}$ = 250 K, $\Delta V$ = 5.0 km s$^{-1}$, see Table~\ref{t:coldens} and Table~\ref{table:column_densities} for remaining parameters), keeping only the column density as a free
parameter. As shown in Fig.~\ref{f:remoca_hoch2ch2cho_ve0_n1s}, 
3-hydroxypropanal is not detected toward Sgr~B2(N1S) and an upper limit column density of $7.9 \times 10^{16} $cm$^{-2}$ was determined. Table~\ref{t:coldens} 
reports this upper limit as well as results obtained 
previously by \citet{sanz-novo2022a} and \citet{Koucky22} for acetaldehyde,
lactaldehyde, 3-hydroxypropenal, propanal, and 2-hydroxyprop-2-enal. 
Table~\ref{t:coldens} indicates that the upper limit derived for 
3-hydroxypropanal is slightly more stringent than the upper limit obtained 
for its structural isomer lactaldehyde by \citet{Koucky22}. We 
find that 3-hydroxypropanal is at least 8.5 times less abundant than 
acetaldehyde in Sgr~B2(N1S). 

\subsection{IRAS 16293-2422B Star-Forming Region}

We utilized data from the ALMA Protostellar Interferometric Line Survey (PILS; \citet{jor16}) to search for 3-hydroxypropanal toward the ``B'' component of the Class 0 protostar IRAS~16293-2422 (IRAS 16293B). The PILS data consist of a full spectral survey of the source between 329.15 and 362.90 GHz at 0.5$''$ angular resolution using ALMA (for details see \citet{jor16}). We examined the spectra toward a position one beam offset from IRAS 16293B that has been the subject of several other studies as part of the PILS program: the position is characterized by bright but narrow emission lines (FWHM of approx 1~km~s$^{-1}$) that makes it ideal to search for new species and provide meaningful upper limits with line blending significantly reduced. Adopting an excitation temperature of 125 K and calculating synthetic spectra for 3-hydroxypropanal, we do not see any indications of lines that can be attributed to the species down to the RMS noise level of the data (4--5~mJy~km~s$^{-1}$). The $3\sigma$ upper limit to the column density for 3-hydroxypropanal derived from these fits comes out to 1.5$\times 10^{14}$~cm$^{-2}$ for the assumed temperature. All assumed source parameters along with the determined upper limit are displayed in Table~\ref{table:column_densities}. The vibrational partition function at 125K was considered when determining this upper limit.

\subsection{NGC 6334I Star-Forming Region}
\label{subsec:ngc}
NGC 6334I is a high-mass star forming region that is host to a variety of complex organic molecules. Recently, several especially large oxygen-bearing species have been detected toward this source \citep{fried_methoxyethanol_2024,mcguire_methoxymethanol_2017}, thus making it a strong target to search for 3-hydroxypropanal. The observational data toward this source was collected using ALMA, with a pointing position focused on the MM1 hot core of this source (J2000 $\alpha$~= 17$^h$20$^m$53.387$^s$ $\delta$~=~-35$^{\circ}$46$^{\prime}$57.533$^{\prime\prime}$). These observations span across regions of ALMA Bands 4, 7, 9, and 10. To ensure uniformity, the data were smoothed using the standard CASA \texttt{imsmooth} routine to achieve a consistent angular resolution of $0.26'' \times 0.26''$ across all frequency bands. The observational details are described in the works of \citet{mcguire_band10_2018}, \citet{mcguire_methoxymethanol_2017}, and \citet{fried_methoxyethanol_2024}. 

\begin{figure}[h!]

\begin{center}
\includegraphics[width=\columnwidth]{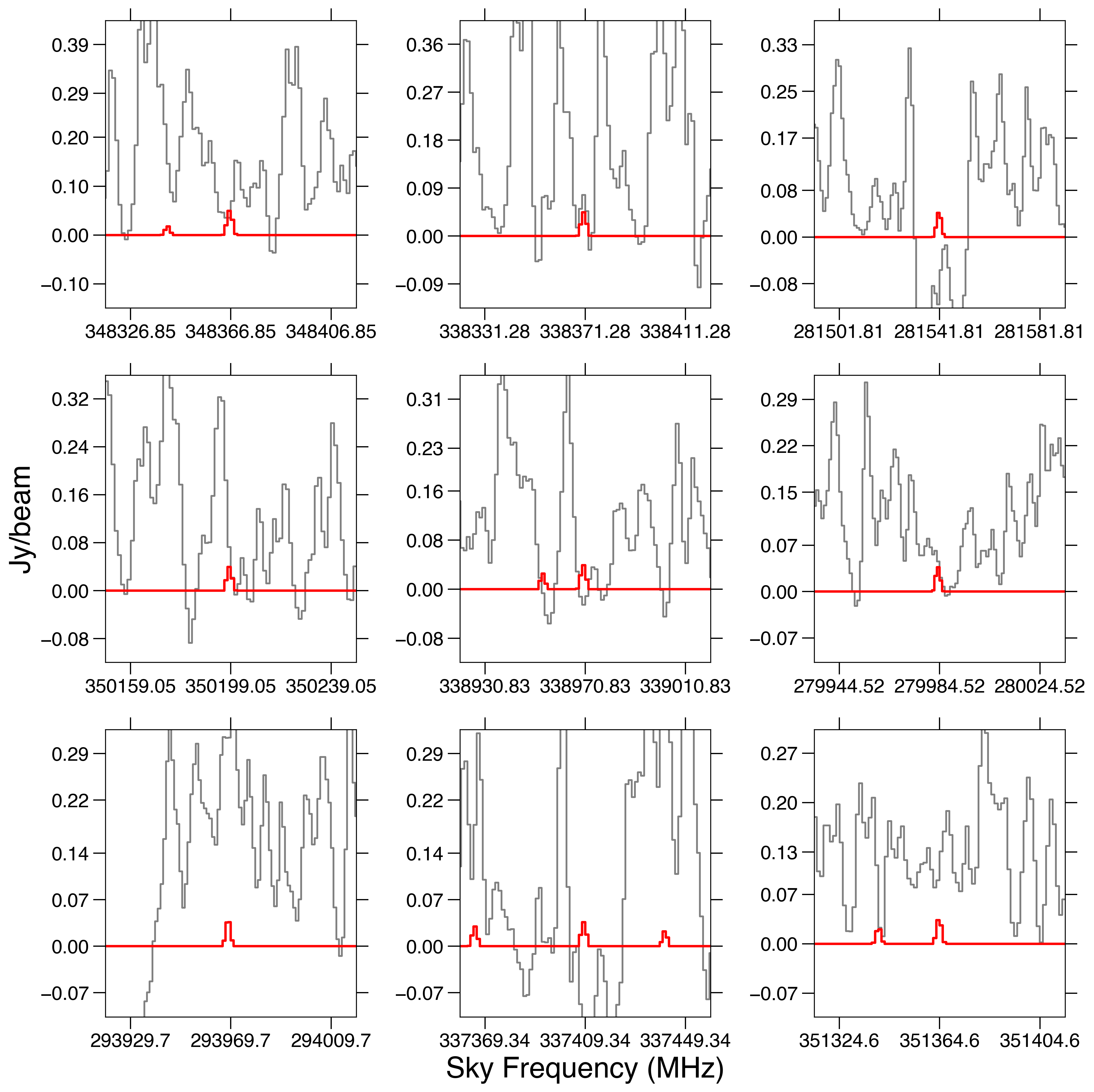}
\caption{
The strongest simulated lines of 3-hydroxypropanal in the observational data toward NGC 6334I. The simulated line profile (red) is overlaid on top of the observational data (gray). The spectrum is simulated at the $3\sigma$ upper limit column density of $4.2 \times 10^{16} $cm$^{-2}$. The other simulation parameters are detailed in the text and Table~\ref{table:column_densities}.}
\label{fig:ngc_ulims}
\end{center}
\end{figure}

\begin{table*}
\caption{Column density of 3-hydroxypropanal in the sources investigated in this work along with the source parameters used in this analysis. The column densities of acetaldehyde and propanal are also noted.}
    \centering
    \makebox[\textwidth]{
    \begin{threeparttable}
    \begin{tabular}{c|cccccccc}
    \hline
     Source & 3-Hydroxypropanal\tnote{a} & $T_{ex}$\tnote{b} & $\Delta V$
     \tnote{c}& $v_{lsr}$&Acetaldehyde & Ratio\tnote{d}& Propanal\tnote{e} & Refs\tnote{f}\\
     & (cm$^{-2}$) & (K) & (km s$^{-1}$) & (km s$^{-1})$ &(cm$^{-2}$) & & (cm$^{-2}$) & \\
    \hline
    G+0.693-0.027 & 8.6$\pm$1.4$\times$10$^{12}$ & 12 & 21.0& 69 & $5.0 \times10^{14}$& 58 &$7.4 \times10^{13}$& 1 \\ 
    Sgr~B2(N) & $\leq7.9 \times10^{16}$& 250 & 5.0& 62 &$6.7 \times 10^{17}$& $\geq$ 8 &$\leq1.6 \times10^{17}$& 1\\
    IRAS 16293-2422B & $\leq1.5 \times10^{14}$& 125& 1.0& 2.6& $1.2 \times10^{17}$&  $\geq$ 800 &$2.2\times10^{15}$& 2,3\\
    NGC 6334I & $\leq4.2 \times10^{16}$& 225& 3.0& -5.2 &$5\times10^{17}$ & $\geq$ 12 & ...& This work\\
    TMC-1 & $\leq 8.6\times 10^{11}$& 7& 0.3& 5.8 & $3.5\times 10^{12}$& $\geq$ 4& $1.9\times 10^{11}$ & 4,5\\
       
    \hline
    \end{tabular}
    \begin{tablenotes}
    \item[a] For Sgr B2(N), IRAS 16293-2422B, and NGC 6334I, the 3-hydroxypropanal column densities are corrected for vibrational and conformational contributions. These contributions are negligable under the cold conditions of G+0.693-0.027 and TMC-1.
    \item[b] Excitation temperature used to determine the column density of 3-hydroxypropanal.
    \item[c] Linewidth used in the simulation to determine the column density of 3-hydroxypropanal.
    \item[d] Ratio of acetaldehyde/3-hydroxypropanal column densities.
    \item[e] A conformer correction to the propanal column density was applied for the Sgr B2(N) and IRAS 16293-2422B star-forming regions. However, the value listed for the cold sources TMC-1 and G+0.693-0.027 correspond to the most stable s-propanal conformer.
    \item[f] References for acetaldehyde and propanal column densities: (1) \citealt{sanz-novo2022a}; (2) \citealt{jorgensen_alma-pils_2018}; (3) \citealt{lykke_alma-pils_2017}; (4) \citealt{cernicharo_discovery_2020}; (5) \citealt{agundez_detection_2023}.

    \end{tablenotes}
    \end{threeparttable}
    }
    \label{table:column_densities}
\end{table*}

The molecular emission was simulated using the \texttt{molsim} Python package \citep{molsim}. This program assumes that the molecular signal is described by a single excitation temperature. The parameters employed for the simulation toward NGC 6334I were an excitation temperature of 225 K, a $v_{lsr}$ of -5.2 km s$^{-1}$, and a linewidth of 3.0 km s$^{-1}$ \citep{Ligterink2020}. The vibrational partition function and conformational correction at 225 K was taken into account when determining the upper limit column density. To establish the upper limit, we identified the 3-hydroxypropanal line in the observations with the highest signal-to-noise ratio under the given simulation parameters that was not blended with the emission of other molecular species. We then calculated the column density necessary for this peak to reach an intensity at least three times the RMS noise level. Since the observations are line confusion limited, we visually determined the RMS noise around the line used to derive the upper limit to be 13 mJy/beam. Ultimately, the 3$\sigma$ column density upper limit was found to be $4.2 \times 10^{16}$ cm$^{-2}$. This upper limit along with the simulation parameters are noted in Table~\ref{table:column_densities}. Figure~\ref{fig:ngc_ulims} displays many of the strongest lines of 3-hydroxypropanal toward NGC 6334I simulated at the upper limit column density. It is clear that molecular emission of 3-hydroxypropanal is not detected at this abundance.

\subsection{TMC-1 Molecular Cloud}

TMC-1 is a cold molecular cloud that has been extensively observed via several broadband line surveys. The detected molecules in this source are most commonly unsaturated carbon-chain species (i.e. \citet{burkhardt_detection_2018, gotham_obs,loomis_stacking_2021}) or cyclic molecules \citep{mcguire_benzonitrile_2018,mcguire_CNP_2021,cernicharo_indene_2021,sita_cyanoindene_2022,loru_ethynylbenzene_2023,cernicharo_ace_2024, wenzel_1pyrene_2024,wenzel_detections_2025,wenzel_coronene_2025}. However, more recently, several large oxygen-bearing fairly saturated species, such as propanal and ethanol, have been observed \citep{agundez_detection_2023}.

\begin{figure}[h!]

\begin{center}
\includegraphics[width=\columnwidth]{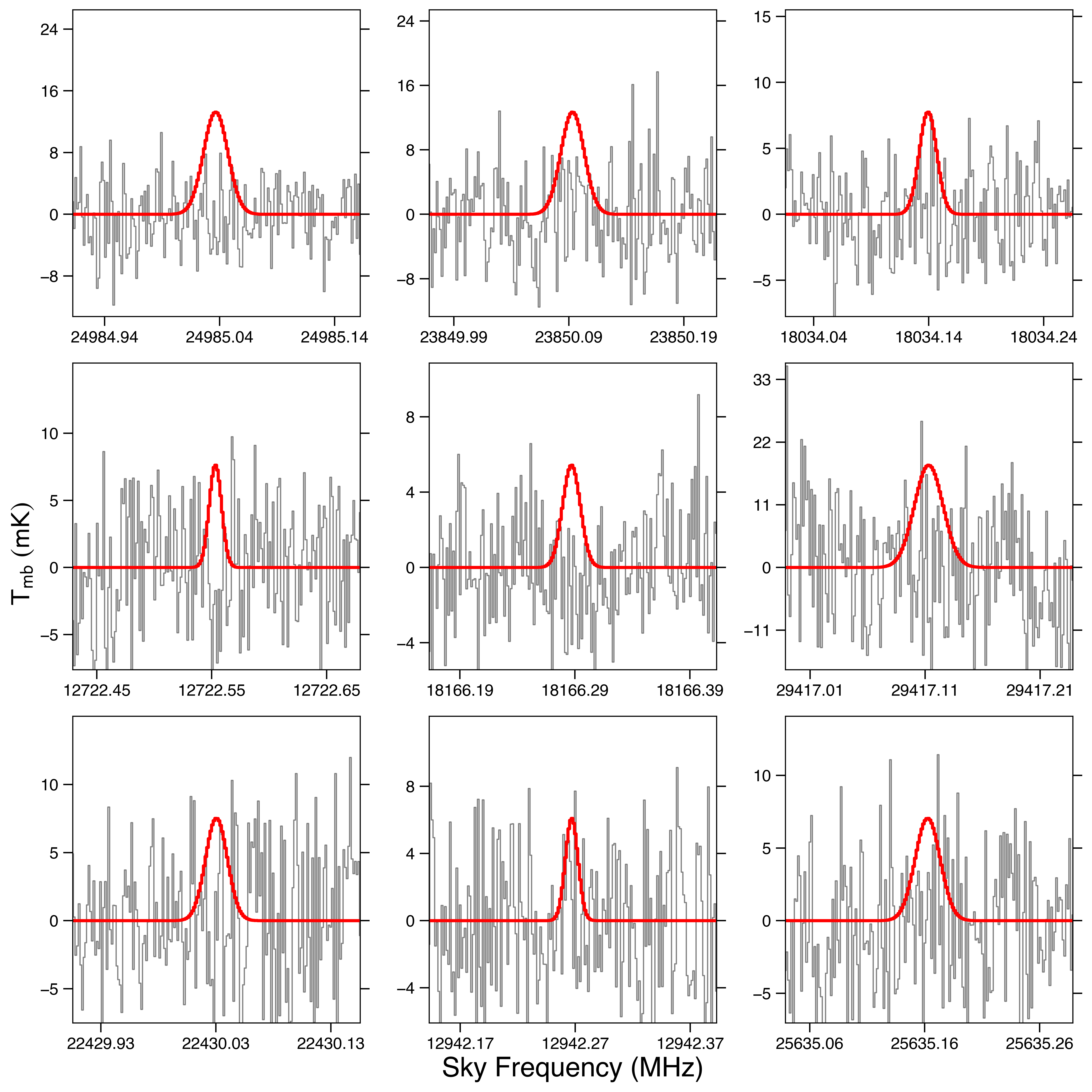}
\caption{
Transitions of 3-hydroxypropanal with the highest signal-to-noise ratio in the GOTHAM DR5 data toward TMC-1. The simulated line profile (red) is overlaid on top of the observational data (gray). The spectrum is simulated at the $3\sigma$ upper limit column density of $8.6 \times 10^{11} $cm$^{-2}$. The other simulation parameters are detailed in the text and Table~\ref{table:column_densities}.}
\label{fig:tmc_ulims}
\end{center}
\end{figure}

We searched for 3-hydroxypropanal in the fifth data reduction (DR5) of the GOTHAM (GBT Observations of TMC-1: Hunting Aromatic Molecules) observations. This line survey was observed using the Robert C. Byrd Green Bank Telescope (GBT). The GOTHAM project is described by \citet{gotham_obs}, while the data reduction process for DR5 will be outlined in a forthcoming paper (Xue et al. in review). This survey has a frequency coverage spanning from 3.9 to 36.4 GHz (with some gaps). The data has an extremely high frequency resolution of 1.43 kHz and a noise level generally around 4-15 mK. The pointing position is focused on the cyanopolyyne peak of TMC-1 at (J2000) $\alpha$~=~04$^h$41$^m$42.5$^s$ $\delta$~=~+25$^{\circ}$41$^{\prime}$26.8$^{\prime\prime}$. The molecular emission was again simulated using \texttt{molsim}, assuming an excitation temperature of 7 K, a $v_{lsr}$ of 5.8 km s$^{-1}$, and a linewidth of 0.3 km s$^{-1}$—parameters that are generally representative of the molecules detected in this source (e.g., \citealt{burkhardt_detection_2018}). The upper limit was determined using the same method as Section~\ref{subsec:ngc}. In summary, the 3-hydroxypropanal transition with the strongest signal-to-noise ratio under the simulation parameters was determined. The column density required for this transition to have an intensity three times the RMS noise level around this line was then derived. The upper limit column density was found to be $8.6 \times 10^{11}$ cm$^{-2}$. Some of the most intense lines of 3-hydroxypropanal toward TMC-1 (simulated at the upper limit column density) are displayed in Figure~\ref{fig:tmc_ulims}.

This, unfortunately, is not an extremely stringent column density upper limit due to the weak transitions of the molecule within the frequency range of the GOTHAM observations. In fact, propanal (which is a less complex species than 3-hydroxypropanal) was observed in the QUIJOTE (Q-band Ultrasensitive Inspection Journey to the Obscure TMC-1 Environment) survey \citep{quijote} of TMC-1 with a column density of $(1.9 \pm 0.7) \times 10^{11}$ cm$^{-2}$ \citep{agundez_detection_2023}. The frequency range of the QUIJOTE survey is 31.0-50.3 GHz. Under the assumed simulation parameters, the strongest simulated line in this range is around twice the intensity of the transition used to derive the upper limit. Moreover, the QUIJOTE survey has a noise level between 0.06 mK at 32 GHz and 0.18 mK at 49.5 GHz \citep{quijote_params}, which is more than ten times lower than the GOTHAM survey. Thus, with additional investigation, this upper limit could likely be further refined toward TMC-1. 

\subsection{Discussion of Chemical Implications}

As mentioned previously, one of the main formation pathways of 3-hydroxypropanal that has been proposed in the literature under astrophysically relevant conditions is the radical recombination of \ce{CH2OH} and \ce{CH2CHO} (Reaction~\ref{eq:radical_recomb}). Although \ce{CH2OH} has not been detected in the interstellar medium, it has been shown experimentally to form from common interstellar precursors, such as hydrogenation of formaldehyde and hydrogen abstraction from methanol \citep{radical_gas,radical_ice,desorption4}. Moreover, as discussed in \citet{wang23}, \ce{CH2CHO} could feasibly be formed via hydrogen abstraction of acetaldehyde (\ce{CH3CHO}). However, such production of \ce{CH2CHO} from acetaldehyde may be inefficient. 

\begin{figure}[h!]

\begin{center}
\includegraphics[width=\columnwidth]{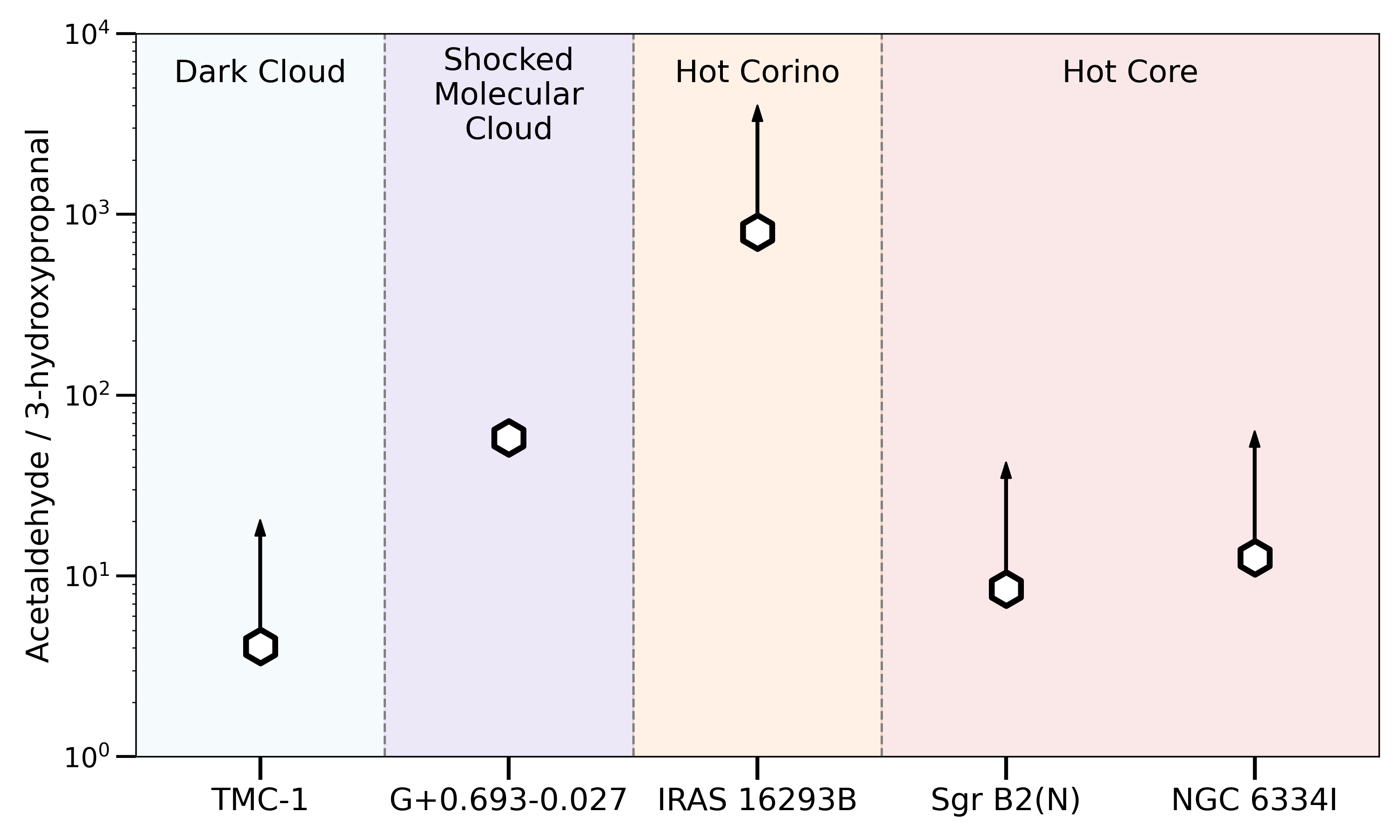}
\caption{
Column density ratio of  acetaldehyde to 3-hydroxypropanal in each of the sources analyzed in this work. The arrows denote an upper limit column density of 3-hydroxypropanal. The values used for this figure and the related references are listed in Table~\ref{table:column_densities}}.
\label{fig:acetaldehyde}
\end{center}
\end{figure}

The C-H bond on the methyl group of acetaldehyde has a higher dissociation energy than the aldehyde hydrogen \citep{berkowitz_three_1994}. Therefore, several gas phase experiments have shown that the \ce{CH3CO} radical is by-far the dominant hydrogen abstraction product of acetaldehyde when reacting with OH \citep{cameron_reaction_2002,butkovskaya_branching_2004}. Moreover, an investigation of interstellar ice analogs found that the predominate initial reaction of acetaldehyde and hydrogen atoms is hydrogen abstraction to form \ce{CH3CO} and \ce{H2} \citep{molpeceres_hydrogenation_2025}. Additionally, during the aforementioned experiments of \citet{wang23}, the \ce{CH3CO} radical was detected in FTIR data of the irradiated methanol-acetaldehyde ices, whereas \ce{CH2CHO} was not directly observed above the detection limit. 

Although it is clear that the production of \ce{CH3CO} is the primary product of the hydrogen abstraction of acetaldehyde, the enhanced cosmic ray ionization rate in G+0.693 may still result in a non-negligible formation of \ce{CH2CHO}. Several previous chemical modeling studies toward G+0.693 have found that a notably enhanced cosmic-ray ionization rate (factor of more than 100 greater than the standard Galactic disk value) is required to reproduce the observed abundances of some molecular cations in this source \citep{Sanz-Novo2024a,Rivilla2022_PO+}. This greater flux of cosmic rays would likely promote the production of radical species in the interstellar ices. These radicals could then recombine on the grain surface, where the surface acts to stabilize the association process and absorb the additional energy. The shocks produced by cloud-cloud collisions in G+0.693 would then cause the molecular inventory stored in interstellar ices to desorb into the gas phase, making them detectable with radio telescopes. 

The column density ratio of acetaldehyde to 3-hydroxypropanal in each of the sources investigated is displayed in Figure~\ref{fig:acetaldehyde}. The lower limit ratio in TMC-1, Sgr B2(N), and NGC 6334I is below the tentatively derived value in G+0.693. Since these are lower limit ratios, it is challenging to draw any conclusions regarding the relative abundances in these three sources compared to G+0.693. For example, the true ratio for TMC-1, Sgr B2(N) and NGC 6334I could in theory be either above or below the tentatively determined value for G+0.693. However, this ratio is notably greater in IRAS 16293B than in G+0.693, thus indicating that there is a clear chemical differentiation in these two sources. One possible explanation is that differences in radiation environments, such as the elevated cosmic ray ionization rate in the Galactic Center, may lead to a higher rate of acetaldehyde dissociation in G+0.693, thereby enhancing the formation of \ce{CH2CHO}. Furthermore, the elevated cosmic ray ionization rate could also result in a higher abundance of atomic hydrogen or carbon, potentially enhancing addition and abstraction reactions on dust grains and promoting the formation of the necessary precursor radical species \citep{requena-torres_galactic_2008,jimenez-serra_observations_2025}. Alternatively, the destruction of 3-hydroxypropanal may be more efficient at the higher temperatures characteristic of the IRAS 16293B hot corino, or it may have been more extensively destroyed during ice processing in the warm-up phase of IRAS 16293B.

Another potential formation route of 3-hydroxypropanal could be the insertion of singlet-excited oxygen atoms (O(\textsuperscript{1}D)) into the C-H bonds of the terminal \ce{CH3} group of propanal (\ce{CH3CH2CHO}). Singlet oxygen has been shown to form quite efficiently via the photolysis of molecules that are known to be abundant in interstellar ices. For example, experimental studies have shown that UV photolysis of \ce{H2O} can produce O(\textsuperscript{1}D) at around a 10\% yield \citep{slanger_photodissociative_1982,stief_flasch_1975}. Moreover, O(\textsuperscript{1}D) is formed at nearly 100\% efficiency following the photolysis of \ce{CO2} at various UV wavelengths \citep{zhu_production_1990,slanger_co2_1971}. Once produced, these excited oxygen atoms can efficiently insert into the bonds of hydrocarbon species \citep{hays13,bergner17,bergner19,daniely_photochemical_2025}. For example, it was found in astrophysical ice analogs that the insertion of O(\textsuperscript{1}D) into the C-H bonds of methane and ethane can form methanol and ethanol with negligible reaction barriers \citep{bergner17,bergner19}. Thus, it may be possible for a similar reaction to occur with propanal on interstellar ices. To the best of our knowledge, among the sources examined in this study, column densities of propanal have been previously reported in the literature for G+0.693 \citep{sanz-novo2022a}, IRAS 16293B \citep{lykke_alma-pils_2017}, and TMC-1 \citep{agundez_detection_2023}. Of note, some propanal transitions were observed in GBT observations  of the Sgr B2(N-LMH) \citep{hollis_green_2004}. However, since the observed lines are mainly in absorption, this indicates that the observations are likely probing the colder, extended gas around the source. On the other hand, only an upper limit column density of propanal has been determined in the ReMoCA observations of Sgr B2(N) analyzed in this paper \citep{sanz-novo2022a}. Additionally, as mentioned previously, the derived column density of propanal from the QUIJOTE survey of TMC-1 is below the upper limit column density of 3-hydroxypropanal that we have derived in the GOTHAM observations; thus, no useful conclusions can be drawn from this result. 

The column density of propanal was found to be $2.2 \times 10^{15}$ cm$^{-2}$ toward IRAS 16293B \citep{lykke_alma-pils_2017}, which would suggest a propanal/3-hydroxypropanal ratio $\geq 14.7$ in this source. On the other hand, as noted in Section~\ref{g693_description}, this ratio has been tentatively determined to be around 9 in G+0.693. The high cosmic ray flux in G+0.693 could result in a strong field of secondary UV photons resulting from the interaction of these cosmic rays with molecular hydrogen \citep{prasad_uv_1983}. This UV-field could therefore promote the production of singlet excited oxygen atoms in the ices in this source, thus resulting in 3-hydroxypropanal formation.

\section{Conclusions}
\label{sec:conclusion}
In this work, we sought to detect 3-hydroxypropanal in the interstellar medium. This molecule shares chemical similarities with several previously detected interstellar species and has been predicted by multiple studies to be a promising detection target. To achieve this, we first synthesized the compound and then measured and analyzed the rotational spectrum of its most stable conformer within the 130–485 GHz range. In the process, we also fitted the rotational spectra of the two lowest lying vibrationally excited states. Due to the relatively small energy gap between these two excited states, some transition frequencies exhibited noticeable perturbations. Thus, we attempted to fit the energy difference between the states along with several Coriolis coupling parameters. Unfortunately, many of the perturbed lines were either extremely weak or below the noise level. Thus, only a small number of perturbed transitions could be incorporated into the fit, which resulted in several of the parameters only being determined with moderate precision. These parameter values could be much improved if better signal-to-noise ratio is achieved for these transitions in future work. 

After analyzing and fitting the spectrum of the ground vibrational state, we searched for this molecule in the radio astronomical observations toward the IRAS 16293-2422B hot corino, the Sagittarius B2(N) and NGC 6334I hot cores, as well as the G+0.693-0.027 and TMC-1 molecular clouds. Molecular emission of 3-hydroxypropanal was tentatively detected for the first time toward G+0.693-0.027 and a column density of (8.6$\pm$1.4)$\times$10$^{12}$ cm$^{-2}$ was derived at an excitation temperature of 12 K. This value suggests that 3-hydroxypropanal is $9\pm2$ times less abundant than propanal in this source.

Unfortunately, no molecular emission of 3-hydroxypropanal was observed in any of the other sources that were investigated in this work. Column density upper limits were therefore determined. We then discussed some potential formation pathways of this molecule, including radical-recombination reactions and oxygen-insertion processes. 

Although a fairly comprehensive number of interstellar sources were analyzed in this work, the measurement of the spectrum across a broadband frequency range and into the submillimeter wave regime now makes it possible to search for 3-hydroxypropanal in various other astronomical objects. One additional potentially strong candidate would be Orion KL, where other \ce{C3H6O2} isomers have been detected \citep{methyl_acetate,peng_alma_2019}. Additionally, The tentative detection of 3-hydroxypropanal reported here prompts the search for other \ce{C3H6O2} isomers toward G+0.693.

\begin{acknowledgments}

\centerline{\textbf{Acknowledgements}}

We would like to sincerely thank the three anonymous referees, who each provided expert criticisms that substantially improved the quality of this manuscript. The authors gratefully acknowledge Dr. Isabelle Kleiner and Dr. Ha Vinh Lam Nguyen for providing additional spectroscopic catalogs for our analysis. This work has been funded by the Czech Science Foundation (GACR, grant No. 24-12586S). Prague authors gratefully acknowledge this financial support. Computational resources were provided by the e-INFRA CZ project (ID:90254), supported by the Ministry of Education, Youth and Sports of the Czech Republic. Z.T.P.F and B.A.M gratefully acknowledge the support of Schmidt Family Futures. B.A.M gratefully acknowledges the support of the Arnold and Mabel Beckman Foundation Beckman Young Investigator Award. B.A.M and C.X gratefully acknowledge support of National Science Foundation grant AST-2205126. V.M.R. acknowledges support through the grant RYC2020-029387-I funded by MICIU/AEI/10.13039/501100011033 and by "ESF, Investing in your future", and from the Consejo Superior de Investigaciones Cient{\'i}ficas (CSIC) and the Centro de Astrobiolog{\'i}a (CAB) through the project 20225AT015 (Proyectos intramurales especiales del CSIC); and from the grant CNS2023-144464 funded by MICIU/AEI/10.13039/501100011033 and by ``European Union NextGenerationEU/PRTR". M.S.N. acknowledges a Juan de la Cierva Postdoctoral Fellowship, project JDC2022-048934-I, funded by MCIN/AEI/10.13039/501100011033 and by the European Union ``NextGenerationEU/PRTR".
I.J.-S., V.M.R. and M.S.-N. acknowledge support from the grant PID2022-136814NB-I00 by the Spanish Ministry of Science, Innovation and Universities/State Agency of Research MICIU/AEI/10.13039/501100011033 and by ERDF, UE. I.J.-S. also acknowledges funding from the European Research Council (ERC) under European Union’s Horizon Europe research and innovation programme OPENS (GA 101125858). The National Radio Astronomy Observatory is a facility of the National Science Foundation operated under cooperative agreement by Associated Universities, Inc. J.-C.G. thanks the Centre National d'Etudes Spatiales (CNES) for a grant. L.M., R.M. and J.-C.G. were supported by the Programme National ``Physique et Chimie du Milieu Interstellaire" (PCMI) of CNRS/INSU with INC/INP co-funded by CEA and CNES

\textbf{Statement of Efforts.}  All authors contributed to the writing and editing of the manuscript.
\end{acknowledgments}
\bibliography{references,library-lucie, hoch2ch2cho_remoca,biblio}
\bibliographystyle{aasjournal}

\appendix

\renewcommand\thefigure{\thesection\arabic{figure}}   
\renewcommand\thetable{\thesection\arabic{table}}    

\setcounter{figure}{0}    
\setcounter{table}{0} 

\section{Complementary Figures}
\label{apendix_a}

\restartappendixnumbering

\begin{center}
\begin{figure*}[h]
     \centerline{\resizebox{1.0
     \hsize}{!}{\includegraphics[angle=0]{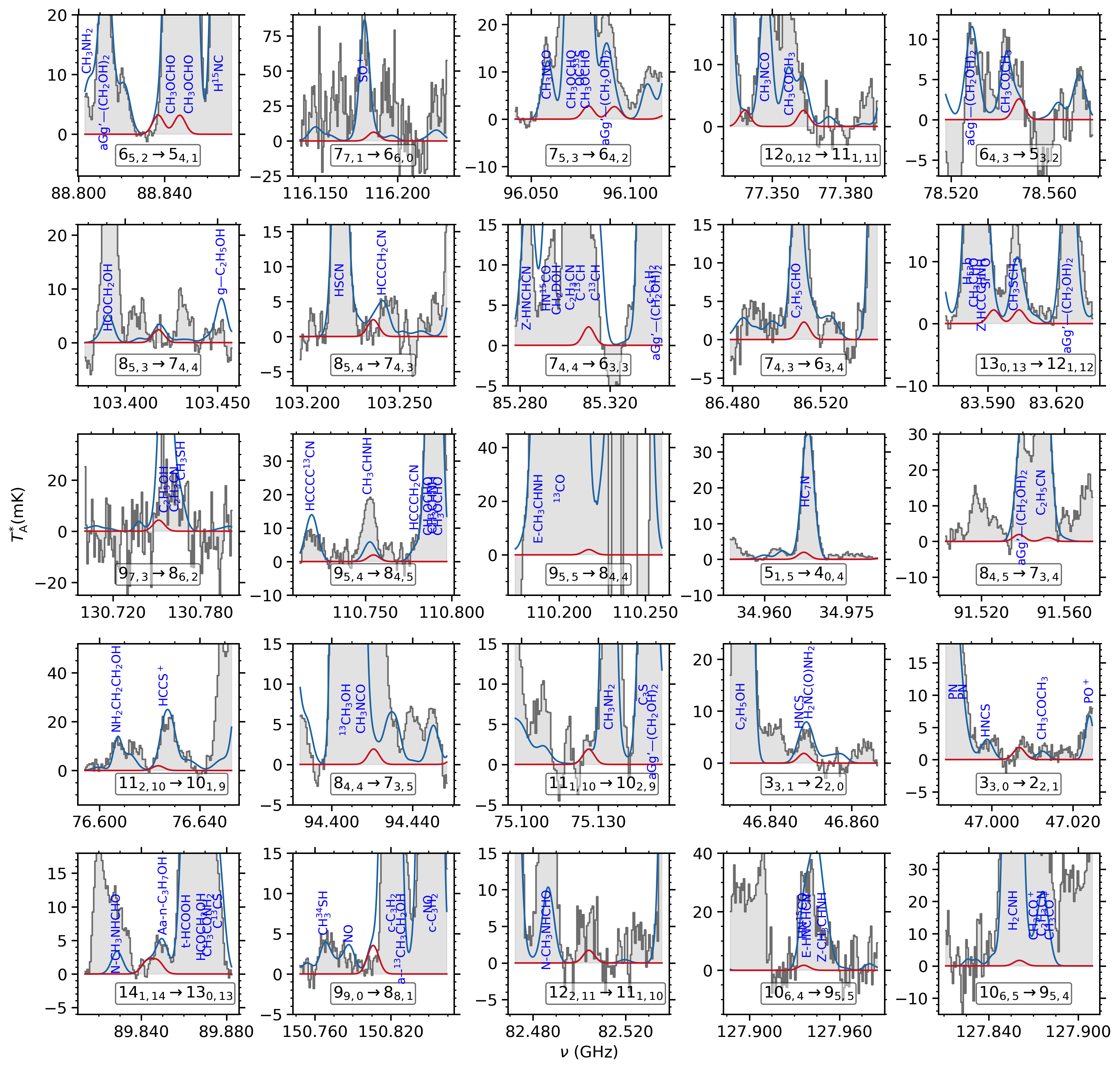}}}
     \caption{Additional transitions of 3-hydroxypropanal not shown in Figure \ref{f:LTEspectrumG0693} which are covered with the survey toward the GC molecular cloud G+0.693–0.027. The result of the best LTE fit of 3-hydroxypropanal is plotted using a red line while the blue line shows the emission from all the molecules identified to date in the survey, including also the target aldehyde (observed spectra shown as gray histograms and light gray shaded area). The quantum numbers involved in each transition using the conventional notation for asymmetric tops are shown in the lower part of each panel. For the panels including multiple lines we provide the quantum number of the lower-in-frequency transition. The transitions are sorted by decreasing intensity.}
\label{f:LTEall1}
\end{figure*}
\end{center}

\begin{center}
\begin{figure*}[h]
     \centerline{\resizebox{1.0
     \hsize}{!}{\includegraphics[angle=0]{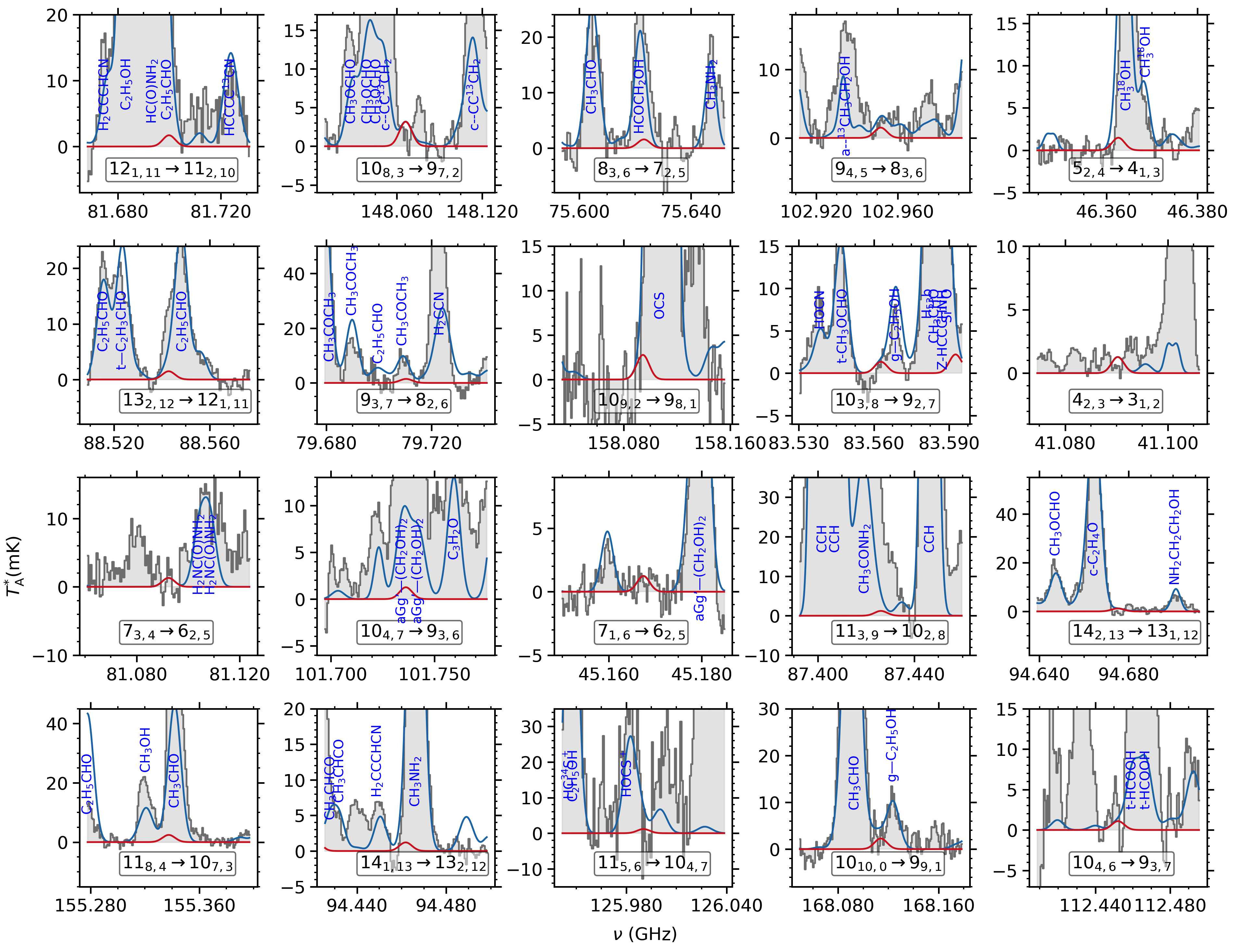}}}
     \caption{Same as in Figure \ref{f:LTEall1} but for additional, weaker lines.}
\label{f:LTEall2}
\end{figure*}
\end{center}

In Figures \ref{f:LTEall1} and \ref{f:LTEall2} we report the remaining brightest transitions of 3-hydroxypropanal that are also covered by the survey conducted toward the molecular cloud G+0.693-0.027 but are not shown in Figure \ref{f:LTEspectrumG0693}, spanning lines down to an intensity comparable to the weakest transition shown in Figure \ref{f:LTEspectrumG0693}. The vast majority of these lines appear heavily blended with the emission from other brighter species previously identified in this survey or fall within a region with lower sensitivity.

\section{Complementary Tables}
\label{appendix:b}
Table~\ref{t:coldens} summarizes key details of the search for 3-hydroxypropanal in Sgr~B2(N1S), including the source parameters such as temperature, source size, and line width, as well as the column densities determined for chemically related molecules.

\begin{table*}[h]
\begin{center}
\caption{ Parameters of our best-fit LTE model of acetaldehyde toward Sgr~B2(N1S) and upper limits for propanal, 3-hydroxypropanal, lactaldehyde, 3-hydroxypropenal, and 2-hydroxyprop-2-enal.}
\label{t:coldens}
\begin{threeparttable}
\setlength{\tabcolsep}{4pt}
\vspace*{-1.2ex}
\begin{tabular}{lcrccrccccr}
\hline\hline

 \multicolumn{1}{c}{Molecule} & \multicolumn{1}{c}{Status\tnote{a}} & \multicolumn{1}{c}{$N_{\rm det}$\tnote{b}} & \multicolumn{1}{c}{Size\tnote{c}} & \multicolumn{1}{c}{$T_{\mathrm{rot}}$\tnote{d}} & \multicolumn{1}{c}{$N$\tnote{e}} & \multicolumn{1}{c}{$F_{\rm vib}$\tnote{f}} & \multicolumn{1}{c}{$F_{\rm conf}$\tnote{g}} & \multicolumn{1}{c}{$\Delta V$\tnote{h}} & \multicolumn{1}{c}{$V_{\mathrm{off}}$\tnote{i}} & \multicolumn{1}{c}{$\frac{N_{\rm ref}}{N}$\tnote{j}} \\ 
  & & & \multicolumn{1}{c}{\small ($''$)} & \multicolumn{1}{c}{\small (K)} & \multicolumn{1}{c}{\small (cm$^{-2}$)} & & & \multicolumn{1}{c}{\small (km~s$^{-1}$)} & \multicolumn{1}{c}{\small (km~s$^{-1}$)} & \\ 
 \hline
 CH$_3$CHO\tnote{k}  $^\star$ & d & 31 &  2.0 &  250 &  6.7 (17) & 1.09 & 1.00 & 5.0 & 0.0 &       1 \\ 
 C$_2$H$_5$CHO\tnote{k} & n & 0 &  2.0 &  250 & $<$  1.6 (17) & 4.46 & 1.18 & 5.0 & 0.0 & $>$     4.3 \\ 
 HOCH$_2$CH$_2$CHO, $v=0$ & n & 0 &  2.0 &  250 & $<$  7.9 (16) & 5.39 & 1.05 & 5.0 & 0.0 & $>$     8.5 \\ 
 CH$_3$CH(OH)CHO, $v=0$\tnote{l} & n & 0 &  2.0 &  250 & $<$  1.7 (17) & 5.68 & 1.01 & 5.0 & 0.0 & $>$     3.9 \\ 
 HOCHCHCHO, $v=0$\tnote{l} & n & 0 &  2.0 &  250 & $<$  7.0 (16) & 2.00 & 1.00 & 5.0 & 0.0 & $>$     9.6 \\ 
 CH$_2$C(OH)CHO, $v=0$\tnote{l} & n & 0 &  2.0 &  250 & $<$  7.2 (16) & 2.41 & 1.00 & 5.0 & 0.0 & $>$     9.3 \\ 
\hline 
 \end{tabular}


\begin{tablenotes}
\item[a] d: detection, n: nondetection.
\item[b] Number of detected lines \citep[conservative estimate, see Sect.~3 of][]{Belloche16}. One line of a given species may mean a group of transitions of that species that are blended together.
\item[c] Source diameter (FWHM).
\item[d] Rotational temperature.
\item[e] Total column density of the molecule. $x$ ($y$) means $x \times 10^y$.
\item[f] Correction factor that was applied to the column density to account for the contribution of vibrationally excited states, in the cases where this contribution was not included in the partition function of the spectroscopic predictions.
\item[g] Correction factor that was applied to the column density to account for the contribution of other conformers in the cases where this contribution could be estimated but was not included in the partition function of the spectroscopic predictions.
\item[h] Linewidth (FWHM).
\item[i] Velocity offset with respect to the assumed systemic velocity of Sgr~B2(N1S), $V_{\mathrm{sys}} = 62$ km~s$^{-1}$.
\item[j] Column density ratio, with $N_{\rm ref}$ the column density of the previous reference species marked with a $\star$.
\item[k] The parameters were derived from the ReMoCA survey by \citet{sanz-novo2022a}.
\item[l] The parameters were derived from the ReMoCA survey by \citet{Koucky22}.

\end{tablenotes}
\end{threeparttable}
 \end{center} 
\end{table*}

\end{document}